# AI-Driven Discovery of High-Temperature Superconductors via Materials Genome Initiative and High-Throughput Screening

**H. Gashmard[1], H. Shakeripour[1*], M. Alaei[1,2]**

*[1] Department of Physics, Isfahan University of Technology, Isfahan 84156-83111*
*[2] Skolkovo Institute of Science and Technology, Bolshoy Boulevard 30, bld. 1, Moscow 121205, Russia*

*[*] hshakeri@iut.ac.ir*

**Abstract**

Inspired by nature, this study employs the Materials Genome Initiative to identify key components of high-temperature superconductors. Integrating artificial intelligence (AI) with high-throughput screening (HTS), we uncover crucial superconducting "genes". Through HTS techniques and advanced machine learning (ML) models, we demonstrate that Functional Convolutional Neural Networks (CNNs) ensure accurate extrapolation of potential compounds. Leveraging extensive datasets from the Inorganic Crystal Structure Database (ICSD), the Materials Project and Crystallography Open Database (COD), our implemented HTS pipeline classifies likely superconductors, with CNN and long short-term memory (LSTM) models predicting transition temperatures and their foundational elements. We address the scarcity of non-superconducting material data by compiling a dataset of 53,196 non-superconducting materials (DataG Non-Sc) and introduce a novel neural network architecture using Functional API for improved prediction, offering a powerful tool for future superconductor discovery. Our findings underscore the transformative potential of combining HTS with AI-driven models in advancing high-temperature superconducting materials, highlighting Pu and H elements (with $T_c$ ~ 100 K) as significant predictors of high-temperature superconductivity, suggesting their role as a crucial "gene" in these materials. Our HTS pipeline predicts 24 new binary Pu-based superconductors with $T_c$ > 50 K at ambient pressure, including 16 compounds exceeding 70 K, in three certain categories, alongside over 40 hydrides with $T_c$ values up to 100 K (21 systems surpassing 64 K).

*Keywords: High-Throughput Screening Pipeline, Materials Genome Initiative, Building Blocks, High-Temperature Superconductors, Artificial Neural Networks, Extrapolation, Convolutional Neural Network, Functional API approach*

## INTRODUCTION

Nature has ceaselessly inspired human ingenuity, guiding the development of novel materials and groundbreaking inventions[1-7]. From the flight of birds inspiring human aviation to the hierarchical structure of bone inspiring materials with enhanced mechanical properties, natural designs have driven countless innovations. For instance, the flexibility and strength of natural fibers have led to the creation of flexible lithium-ion batteries, while artificial photosynthesis replicates natural processes to capture solar energy effectively. Nature-inspired morphologies have also improved supercapacitor performance, and materials derived from spider silk offer sustainable alternatives to plastics. Innovations like Velcro, inspired by burdock plants, and bio-inspired water purification systems demonstrate further the broad spectrum of applications drawn from nature's designs[1-8].

A profound inspiration from nature is the concept of the "genome", which revolutionized our understanding and catalyzed significant scientific progress. The Human Genome Project (HGP), which decoded the human genome, marked a pivotal milestone in medical and biological research by uncovering the fundamental building blocks of life[9-13]. Similarly, in materials science, the concept of a "gene" can be likened to its biological counterpart, a fundamental unit responsible for imparting unique properties to a material. For example, the oxide planes enabling superconductivity in cuprate materials can be considered a "gene." In this context, a material's gene might consist of an element, a combination of elements, or a unique structural feature that defines key properties within a group of materials.

The Materials Genome Initiative (MGI), launched in 2011, aims to accelerate the discovery, design, and deployment of new materials by integrating data-driven approaches, computational tools, and experimental techniques. Central to this initiative is the creation of extensive materials libraries and the use of high-throughput synthesis and characterization to rapidly screen materials, significantly reducing the time and cost of research. ML models, particularly suited for high-throughput screening (HTS), are increasingly being employed to predict materials properties, further accelerating the discovery of advanced materials. Inspired by the transformative impact of high-throughput methods in drug discovery, this approach is now gaining traction in superconductivity research, with applications in high-throughput synthesis, ab-initio calculations, and ML-driven experimentation[11-20].

This article highlights the potential of ML approaches to identify the "genes" or fundamental building blocks of materials, contributing significantly to the MGI. A key challenge in this field is accurately identifying these material genes, determining which component of a material should be designated as the "gene" within a specific category. However, evaluating a broader range of materials enhances the precision of this determination. Just as the human genome contains about 20,000 genes, the material genome comprises numerous genes, each governing distinct material properties[11,21-23]. Here, we demonstrate the discovery of new genes for high-temperature superconductors (HTSCs) through HTS and advanced ML models.

Superconductivity, a macroscopic manifestation of quantum phenomena, arises from the finite attraction between paired electrons[24,25]. The performance of superconductors is defined by their critical temperature ($T_c$) and critical current density ($J_c$), which determine their suitability for various applications, particularly in electrical engineering and energy systems[26,27]. A major limitation of superconducting materials is their low transition temperature, which necessitates costly cooling systems to maintain operational states. This challenge has driven the scientific community to prioritize the discovery of materials with higher transition temperatures, which would reduce costs and enable broader applications. However, the strong electron-electron correlations in these materials make first-principles calculations to determine their electronic structures or predict $T_c$ values extremely challenging[28-30], necessitating alternative discovery methods.

To efficiently discover new superconducting materials, HTS methods are essential. HTS enables researchers to evaluate numerous material candidates rapidly, significantly accelerating the discovery process. Emerging technologies, such as artificial intelligence (AI), ML algorithms and data-driven modeling, hold the potential to significantly reduce the time and expense involved, making the search for novel superconducting materials more efficient and cost-effective[16-19]. Materials science has entered its fourth stage of evolution, referred to as "data-driven materials science."[20,31-33]. By leveraging vast datasets and cutting-edge AI algorithms, supported by high-speed data processing hardware, researchers can vastly expand their capacity to explore and discover new materials[16,17,20,31,33].

ML focuses on developing algorithms that improve through experience by identifying patterns, trends and insights in data, enabling prediction and analysis. Data inherently contains knowledge, regularities, and patterns, which ML algorithms extract to create predictive models. In materials science, these data-driven approaches accelerate the discovery of new materials by uncovering hidden patterns and suggesting potential candidates from existing datasets[20,34-42].

Stanev *et al.* (2018) introduced an ML approach to predict the $T_c$ using data from the SuperCon database, which includes over 12,000 known superconductors. By employing classification and regression models (Random Forest) based on coarse-grained chemical compositions, the authors achieved approximately 92% accuracy in determining whether materials exhibit $T_c$ values above or below 10 K. They also integrated these models into a single pipeline to screen the entire Inorganic Crystallographic Structure Database (ICSD), comprising around 110,000 materials, and identified more than 30 new candidate superconductors[24]. Roter and Dordevic (2020) developed an unsupervised ML model to predict the $T_c$ of superconductors using only their chemical compositions from the SuperCon. Their model achieved a coefficient of determination ($R^2$) of approximately 0.93 and a root-mean-square error (RMSE) of about 8.91 K. To enhance the dataset, they incorporated around 3,000 non-superconducting compounds, including insulators and semiconductors. Using Singular Value Decomposition (SVD) and k-Nearest Neighbors (KNN) for classification, they achieved 96.5% accuracy and identified several new superconductors[43]. Zeng *et al.* (2019) proposed an innovative atom table convolutional neural network (ATCNN) model to predict material properties, including $T_c$, band gap ($E_g$), and formation energy ($E_f$), using only their chemical compositions. The model distinguished between superconducting and non-superconducting materials by incorporating 9,399 stable insulators with a DFT band gap greater than 0.1 eV into the dataset of 13,598 superconductors. These insulators, sourced from the Materials Project, were classified as non-superconductors. The ATCNN model achieved an RMSE of 8.19 K and an $R^2$ of 0.97. Additionally, the framework was used for HTS, leading to the identification of 20 new superconductors[19]. Konno *et al*. (2021) employed a deep learning model to predict the $T_c$ of superconductors using SuperCon database, achieving an $R^2$ value of 0.92. They introduced two innovative methods: "reading the periodic table," which enables the model to learn elemental properties, and "garbage-in," which generates synthetic data for non-superconductors by assuming that most inorganic materials in the Crystallography Open Database (COD) do not exhibit superconductivity with finite $T_c$ values[28]. Quinn and McQueen (2022) trained convolutional neural networks (CNNs) to identify and predict high-$T_c$ superconductors, achieving over 95% accuracy in classification models and an $R^2$ greater than 0.92 in regression models. They developed a pipeline combining these models to analyze approximately 130,000 crystal structures from the Materials Project, identifying several candidate materials, including infinite-layer nickelates, with predicted $T_c$ exceeding 30 K [29]. Pereti *et al*. (2023) introduce an ML method utilizing DeepSet technology to classify and predict the $T_c$ of superconductors based on their chemical composition. This method performs both classification and regression tasks, distinguishing superconductors from non-superconductors and predicting their $T_c$. Trained on datasets from the SuperCon database, the model achieved an RMSE of 9.5 K and an $R^2$ value of 0.92. To address the rarity of superconductors, all materials in the COD were labeled as non-superconductors[44].

The following two studies have explored *extrapolation* in superconducting materials. Moscato *et al*. (2023) used a dataset of 21,263 superconductors from Hamidieh[45] to predict $T_c$ based on their physical-chemical characteristics. For extrapolation, they trained their model on the bottom 90% of the data ($T_c < 89$ K) and tested it on the top 10% ($T_c \geq 89$ K), achieving the extrapolation error with an RMSE of 36.3 K using their Spline Continued Fraction Regression (Spln-CFR) model, which outperformed XGBoost (RMSE = 37.3 K) and Random Forest (RMSE = 38.1 K). In terms of interpolation, XGBoost performed best with an RMSE of 9.47 K, while Spln-CFR had an RMSE of 10.99 K[46]. Meredig *et al*. (2018) investigated the effectiveness of ML in discovering high-$T_c$ superconductors. They introduced two innovative techniques: leave-one-cluster-out cross-validation (LOCO CV) and a nearest-neighbor benchmark, which provide a more realistic assessment of model performance by addressing the challenges of extrapolation in highly clustered training data. Their findings suggest that ML-guided iterative experimentation could outperform standard HTS in identifying breakthrough materials[18].

This study presents a comprehensive analysis of *ten* traditional ML and *nine* neural network models, focusing on their extrapolation capabilities. Our results indicate that only a limited subset of neural network-based models can effectively extrapolate, with the majority lacking this feature. Additionally, our analysis reveals that approximately 96% of known superconducting materials do not exhibit a band gap, prompting us to compile a new dataset specifically for non-superconducting materials. Using this dataset, we developed classification models, employing ensemble learning techniques to enhance predictive robustness and reliability. Furthermore, we developed models to predict the transition temperatures of superconducting materials, leading to the identification of high-transition-temperature superconductors. By combining HTS methods with advanced ML approaches, we identified several compounds with promising superconducting potential, particularly those containing Pu and H elements.

## RESULTS AND DISCUSSION

### High-Throughput Screening for High-Temperature Superconductor Discovery within the Materials Genome Initiative Framework Utilizing Machine Learning Models

In recent years, some studies in the field of superconducting materials have referenced the term MGI without substantively employing its framework or principles[11,12,14,15]. This study explores the "materials genes" associated with high-$T_c$ superconducting materials through HTS of large datasets, including the ICSD, the COD and the Materials Project. By employing Artificial Neural Network (ANN) models, we identified several compounds with potential superconducting properties, characterized by high $T_c$. Notably, the majority of these predicted compounds contained Plutonium and Hydrogen, suggesting that the neural networks recognized Pu and H as significant "material genes" linked to elevated superconducting transition temperatures. These findings highlight the potential role of these elements in the design and discovery of new high-$T_c$ superconducting materials. A flow chart for predicting high-$T_c$ superconductors is shown in Figure 1.

### 1. Evaluating Extrapolation Capabilities of Machine Learning Models for High-Throughput Screening in Superconductor Discovery

Computational tools for HTS of large datasets often rely on ML models. A critical aspect of this process is identifying which ML algorithms possess extrapolation capabilities, as our objective is to discover superconducting materials with high $T_c$. In this study, we conducted a comprehensive evaluation of both traditional ML algorithms and advanced ANN models to determine their extrapolation capabilities. This addresses a significant gap in previous studies, particularly the insufficient focus on the extrapolation potential of ML models in HTS of large datasets, which is essential for accurately predicting material properties beyond the range of existing experimental data. By emphasizing the importance of extrapolation, our work enhances the effectiveness of HTS processes and improves the discovery of new materials with desirable properties.

### 1.1. Extrapolation and Interpolation in Machine Learning

Extrapolation and interpolation are fundamental concepts in ML, particularly within the context of supervised learning algorithms. Interpolation involves predicting outcomes for new observations within the known sample space of the training data. In contrast, extrapolation refers to predicting outcomes beyond the range of the training data, which is inherently more challenging due to the lack of known data points in the target region[47,48].

This study aims to identify and evaluate the most effective ML models that demonstrate both extrapolation and interpolation capabilities for analyzing datasets of superconducting materials. A suitable model is one that can accurately perform interpolation within the known data range and reliably extrapolate beyond it.

### 1.2 Comprehensive Evaluation of Extrapolation Capabilities in Traditional ML and Artificial Neural Network-Based Models

In this section, we present and analyze the results of a comprehensive evaluation of various traditional ML algorithms (10 models) and ANN-based models (9 models) to assess their extrapolation capabilities, focusing on their performance and suitability for HTS.

To evaluate, we utilized the DataG dataset[33]. It includes 13,022 superconducting materials with $T_c$ ranging from 0.01 K to 135.8 K. For the evaluation, we split the dataset into training and testing subsets: 95% of the data (12,371 samples with $T_c$ between 0.01 K and 88 K) was allocated to the training set, and 5% (651 samples with $T_c$ between 88 K and 135.8 K) was reserved for testing. We trained the models on the training set and evaluated their performance on the test set.

To quantify the performance and extrapolation capabilities of the models, we calculated RMSE and Mean Absolute Error (MEA) for each model. This methodology not only identifies models capable of generalizing unseen data but also provides insights into their ability to predict the properties of superconducting materials beyond the training conditions.

**1.2.A. Extrapolation Capability in Traditional Machine Learning Models**

In this section, we evaluate the most prominent traditional ML algorithms to assess their effectiveness in extrapolating the $T_c$ of superconducting materials. This evaluation focuses on identifying models capable of accurately predicting outcomes beyond the range of the training data.

Figure 2a illustrates the RMSE associated with the performance of ten traditional ML models. Notably, following the optimization of various hyperparameters for each model, the XGBoost and CatBoost models demonstrated superior performance, yielding the lowest error values. However, their error rates suggest limitations in extrapolation capabilities. This indicates that while XGBoost and CatBoost may outperform others in predictive accuracy within the dataset's range, their ability to generalize beyond the trained $T_c$ range remains constrained. Figure 2b illustrates the MAE, used as an additional evaluation criterion to quantify the extrapolation error rate specifically for the upper 5% of data entries (651 materials with the highest $T_c$ in the dataset) when employing traditional ML algorithms.

Figure 3 provides an analysis of the number of compounds and the percentage of successful $T_c$ extrapolation within the test dataset. For instance, XGBoost model extrapolated the transition temperatures for 171 out of the 651 test substances, equating to a success rate of 26.28%. Despite this, the XGBoost exhibited the highest extrapolation rate among traditional ML models. Nevertheless, this level of extrapolation underscores the limited predictive power of traditional ML algorithms in accurately identifying high-$T_c$ superconductors.

**1.2.B. Extrapolation Capability in Artificial Neural Network-Based Models**

Artificial Neural Networks (ANNs), inspired by the architecture and functionality of the human brain [8,49-52], are nonlinear statistical models designed to mimic the function of biological neural networks, enabling complex data processing and pattern recognition[53]. Recently, ANNs have become widely utilized and effective tools for tasks such as regression, classification, clustering, pattern recognition, and prediction across numerous fields[52].

Although ANNs have emerged as powerful tools for predicting materials properties, their ability to extrapolate beyond known data distributions remains a critical concern. Here, we employ ANNs to conduct both classification and regression analyses on superconducting materials, evaluating their extrapolation capability, like the approach taken with traditional ML models.

We evaluated the error metrics associated with each ANN model. Figure 4a illustrates the relative RMSE for various models developed using both the Sequential and Functional APIs. The results indicate that the CNN models, whether developed using the Sequential or Functional approach, exhibited the lowest error rates during optimization. Additionally, a Functional API model integrating Long Short-Term Memory (LSTM) and Gated Recurrent Unit (GRU) structures demonstrated the least error in extrapolation. Notably, as shown in Figure 4a, the CNN model achieved the lowest error among both traditional ML models and neural network-based approaches in predicting $T_c$, representing a significant achievement in this study. Figure 4b presents the MAE for the upper 5% of data entries, comprising 651 materials with the highest $T_c$ in the dataset.

Figure 5 analyzes the number of compounds and the percentage of successful $T_c$ extrapolations within the test dataset. Notably, the CNN Functional API model extrapolated $T_c$ for 529 out of the 651 test samples, achieving an accuracy rate of 81.26%. These findings suggest that CNN models offer a promising approach for identifying new superconducting materials with high $T_c$.

Among all traditional ML and neural network-based models, see Figure 2 and 4, the CNN model consistently exhibited the lowest error in extrapolating $T_c$, highlighting its superior predictive accuracy for extrapolation tasks in this domain.

### 2.2 Enhancing Classification Models through Ensemble Learning: Integrating Traditional Machine Learning and Artificial Neural Networks

The design and implementation of an effective classification model are critical components of the HTS pipeline for discovering new superconductors. To enhance and ensure the predictive accuracy of our models, we applied the concept of Ensemble Learning, which provides a robust framework for improving classification performance.

We employed both Magpie[59] and Jabir[33] atomic descriptors to generate the feature space for materials. By using these two distinct feature spaces, our aim was to compare the prediction accuracy of the classification models associated with each feature space independently.

#### 2.2.A. Development of Traditional Machine Learning Models for Classifying Superconducting and Non-Superconducting Materials

In this section, we assess various traditional ML models to compare their performance in classifying superconducting and non-superconducting materials. In the preceding 10-fold cross-validation step, the optimal set of hyperparameters was identified by selecting the combination that yielded the highest average performance across all folds. Once the traditional ML models were trained using the feature space generated by the Jabir package, the evaluation metrics were computed, and the results are presented in Table 1. As evident from the table, the CatBoost model outperforms other traditional models in classification tasks. The performance of the traditional ML models utilizing Magpie descriptors was also evaluated. It was observed that the models generally performed better with features derived from the Jabir package. Therefore, the results related to the performance of the models with Magpie descriptors are not presented in this section.

Figure 6 displays the confusion matrices for the CatBoost model and for 10% test data. Figure S10 presents the ROC curve for the CatBoost model. A higher AUC value indicates better overall performance.

**2.2.B. Development of Artificial Neural Network Models for Classifying Superconducting and Non-Superconducting Materials**

We employed two specific models, CNN and MLP, to develop classification models for distinguishing superconducting materials from non-superconducting materials. Given their promising performance, we did not find it necessary to evaluate additional neural network-based models. We assessed the performance of the models using confusion matrices, which provide a comprehensive overview of their classification performance.

To ensure the reliability of our results, we performed a 10-fold cross-validation on the classification models. This approach divides the dataset into ten subsets, with evaluation metrics averaged across all folds to ensure consistent and reliable performance. The results, summarized in Tables 2.

For final evaluation of model performance, the dataset was divided into a training set (90%) and a testing set (10%). After training the models, the evaluation metrics were calculated, as shown in Table 3. Figure 7 illustrates the confusion matrices for the CNN model. Figure S11 presents the ROC curve for the CNN model, illustrating its performance in distinguishing between superconducting and non-superconducting materials. As can be seen from Figs. 7 and S11, and by comparison with Fig. 5, the confusion matrices of traditional model, utilizing Jabir features yield more reliable results.

**2.2.C. Integration of Four Advanced Classification Models via Ensemble Learning**

To leverage the advantages of ensemble learning, we selected two of the most effective traditional ML models, XGBoost and CatBoost, and two high-performing neural network models, CNN and MLP. By employing these four models, we aimed to accurately classify materials as either superconductors or non-superconductors. We applied the ensemble approach to enhance our confidence in predictions. A material is classified as superconducting only if at least two of the four models independently identify it as such.

The application of ensemble learning for classifying superconducting materials represents a novel approach in this study, with potential applications across various fields in materials science.

**2.3 Predicting Transition Temperatures of Superconducting Materials: A Comparative Analysis Using Traditional Machine Learning and Neural Network Models**

Following the successful implementation of classification models for identifying superconducting materials, the next critical step in our research pipeline centers on predicting the $T_c$ of these materials. We believe that an effective and robust model for discovering novel high-temperature superconductors must excel in both accurate interpolation and extrapolation. To this end, we developed five distinct ML models, two based on traditional ML techniques (CatBoost and XGBoost) and three ANN architectures (CNN, LSTM, and MLP), which were subsequently compared to evaluating their performance. This diverse selection enabled a thorough evaluation of various predictive approaches, allowing us to identify the most effective models for accurate $T_c$ prediction.

For the CatBoost and XGBoost models, we utilized the DataG dataset. Feature selection was performed using the Jabir package. After feature selection, we trained the models, allocating 90% of the dataset for training and the remaining 10% for testing to assess model generalizability. The performance metrics ($R^2$) of these models are illustrated in Figure 8a. While CatBoost and XGBoost models demonstrated strong interpolation abilities for predicting the $T_c$ of superconducting materials, neural network-based models, particularly CNNs, exhibit superior extrapolation performance, as highlighted in Section 1.2.B. Consequently, neural network models are essential at this stage of the research pipeline to achieve accurate $T_c$ prediction. To address this, we designed a unique architecture for three neural network models: CNN, LSTM, and MLP.

**2.3.A Developing a Novel Architecture for Artificial Neural Networks to Predict Transition Temperatures in Superconducting Materials**

Given the limitations of traditional ML models in extrapolation, we turned to ANNs, which excel at learning complex nonlinear relationships and generalizing from training data to unseen scenarios. For training the models, we utilized the DataG dataset and generated a relatively comprehensive feature space with the Jabir and Magpie tools. Specifically, we integrated all 322 features from the Jabir package with 132 features from Magpie. Using the Soraya Python package, we selected 37 of the most significant features to input into the models during the intermediate stage of model development.

The neural network architecture utilized in this study was designed using the Functional API framework. As discussed in Section 1.2.B, the Functional API offers enhanced flexibility and additional capabilities. The architecture is designed to input the 322 features from the Jabir package and the 132 features from the Magpie package separately and in parallel. After processing through various layers, these two streams of data are merged. At this stage of integration, an additional 37 selected features, identified using the Soraya package, are incorporated into the network. This unique Functional API architecture allows for the independent processing of various feature sets, enabling the separate entry of feature space information into the network for enhanced predictive capacity. This approach facilitates the retrieval and utilization of critical information at any stage within the network. For instance, in CNNs, information initially enters the convolutional layers and undergoes transformations. In this architecture, information pertaining to the 37 most crucial features is injected into the neuronal layers after the convolutional layers. This integration method maximizes the influence of essential features, enhancing model performance and accuracy in predictive tasks. Figure 8a illustrates the performance of the neural network models based on the $R^2$ evaluation metric, while Figure 8b shows the RMSE for various models developed for predicting the $T_c$ of superconducting materials. The CatBoost model achieved the lowest RMSE (6.86 K) among traditional ML models, while the CNN model outperforms other neural networks with an RMSE of 7.96 K.

### 2.4 High Throughput Screening of Large Databases for High-Temperature Superconductor Discovery and Building Blocks Identification

In prior steps, we developed robust classification models to identify superconducting materials through comprehensive research on various traditional ML models and ANNs. These were evaluated for their ability to extrapolate $T_c$. Following this, we developed regression models to predict $T_c$ using the best extrapolative models, specifically CNNs and LSTM networks. These models were integrated into a HTS pipeline to systematically analyze large material databases (ICSD, Materials Project, COD), enabling the identification of high-temperature superconductors and their fundamental building blocks ("genes"). As can be seen in Table 4, our analysis revealed *two* distinct families of superconductors, with Pu and H identified as the elemental genes governing their superconducting properties. Predicted compounds within these families exhibit maximum $T_c$ values of ≈ 100 K (Table 4), all characterized by a zero band gap, a hallmark of superconducting potential.

Notably, while materials containing Pu and H were initially observed to display elevated $T_c$ values during screening, systematic validation confirmed Pu and H as the pivotal genes for high-temperature superconductivity. Experimental benchmarks for Pu-containing superconductors, such as $PuCoGa_5$[54], report a maximum $T_c$ of 18 K, the highest previously documented for this class. Strikingly, here, it seems our models raised this $T_c$ to nearly 100 K for $Pu_{28}Zr$ or 89 K for $Pu_2Co_{17}$ compounds, for instance.

To the best of our knowledge, this work pioneers the application of the biologically inspired "gene" concept to superconductors, identifying elemental building blocks critical to high-$T_c$ behavior. This approach marks a unique intersection of materials science and AI, offering a novel framework for understanding and discovering superconducting materials.

#### 2.4.A Plutonium-Based Superconductors

The first family under investigation comprises Pu-based compounds (see Table 4), exemplified by $Pu_{28}Zr$ with $T_c \approx 100$ K, whose crystallographic configuration is depicted in Figure S1. Plutonium (Z=94), an actinide element, exhibits a balance between localized and itinerant 5f electronic states, leading to strong electron correlations and hybridization with Zr conduction electrons. This interaction enhances quasiparticle density at the Fermi level, influencing superconducting pairing.

Unlike conventional phonon-mediated superconductivity, pairing in plutonium-based systems appears to be driven by magnetic fluctuations amplified by strong spin-orbit coupling and mixed valence states[54].

Plutonium, the sixth actinide, exhibits unparalleled electronic complexity due to its intermediate 5f electron localization, positioned between itinerant early actinides (Ac–Np) and localized late counterparts (Am–No)[55,56]. This dual behavior, sensitive to structural and external perturbations (e.g., temperature, doping), drives unique bonding and correlated phenomena. Early actinides display itinerant 5f electrons akin to transition-metal d-orbitals, forming narrow bands with high Fermi-level density of states (DOS) and suppressed local moments[57,58]. In contrast, late actinides adopt localized 4f-like configurations. Pu's mixed valence and strong spin-orbit coupling enable multiple atomic radii, favoring dense liquid packing over crystalline symmetry (e.g., bcc, fcc) due to strain destabilization[56]. Brewer[59] further predict limited Pu solubility with alkali/alkaline earth metals but enhanced compatibility with 3d transition metals, reflecting its multiconfigurational electronic ground states.

Experimental progress on Pu-based superconductors began with the discovery of heavy-fermion superconductivity in $PuCoGa_5$ ($T_c \approx 18$ K) and $PuRhGa_5$ ($T_c \approx 8.5$ K)[60] in the early 2000s. These materials, crystallizing in the tetragonal $HoCoGa_5$ structure, exhibit spin-fluctuation-mediated superconductivity, likely rooted in proximity to a magnetic quantum critical point. Their relatively high $T_c$ for heavy-fermion systems highlights the role of Pu's 5f electrons in generating strong electronic correlations and hybridized states. To date, no ambient-pressure Pu-based superconductors with $T_c$ exceeding 20 K have been experimentally confirmed.

Theoretical efforts have focused on leveraging Pu's complex electronic structure to predict new superconducting phases. Density functional theory (DFT) and dynamical mean-field theory (DMFT) studies emphasize that Pu's 5f orbitals, poised between localized and delocalized behavior, create fertile ground for unconventional superconductivity. Prior predictions include hypothetical Pu-H compounds under high pressure, but none have yet surpassed the $T_c$ of known actinide superconductors[61-63].

Here, we predict 24 new binary Pu-based superconductors with $T_c > 50$ K at ambient pressure, including 16 compounds with $T_c > 70$ K (see Table 4). In Table 4, three categories of predicted Pu-based superconductors are identified: (i) Pu combined with transition elements such as Y, Zr, Os, Co and Mn. (ii) $Pu_5X_3$ compounds (X=Ir, Ru, Os) which Pu combined with three transition elements and (iii) the $Pu_3X'$ series (X'=Yb, La, Dy, Ce, Th, Nd, Er, Pm, Tm, Ho, Sm and Tb), combined with 11 Lanthanide group elements, with $T_c$ values ranging from 65 K ($Pu_3Tb$) to 83 K ($Pu_3Yb$). Additionally, a compound containing Pu and K is also predicted.

The elevated $T_c$ in $Pu_3X'$ compounds, for instance, likely stems from the interplay of three key factors: a) Strong electronic correlations and hybridization: The hybridization of Pu-5f electrons with X'element (e.g., Yb-4f, La-5d) orbitals generates heavy quasiparticles and enhances the density of states near the Fermi level. This hybridization is modulated by the ionic radius of X': smaller ions (e.g., $Yb^{3+}$) impose chemical pressure, shortening Pu-Pu distances and amplifying 5f-electron itinerancy, while larger ions (e.g., $La^{3+}$) may stabilize competing magnetic fluctuations. b) Spin-orbit coupling (SOC): Pu's strong SOC splits 5f states into narrow bands, potentially creating van Hove singularities or flat bands that boost pairing interactions. c) Valence and magnetic tuning: elements like Ce (mixed $4f^0/4f^1$) and Dy (localized 4f moments) introduce valence fluctuations or magnetic exchange, which could suppress competing orders (e.g., magnetism) and stabilize superconducting pairing.

These predictions highlight the untapped potential of engineered plutonium-based superconductors, particularly in ambient-pressure conditions. If experimentally validated, these materials could: Provide new insights into f-electron pairing mechanisms in high-$T_c$ systems, advance our understanding of unconventional superconductivity in actinides and open pathways for tuning $T_c$ via chemical substitutions or pressure-induced modifications.

Further examination of other predicted Pu-containing compounds reveals additional structural details. See Supplementary Information.

### 2.4.B Hydride Superconductors: Toward Ambient-Pressure Stability

The second family of interest consists of H-based compounds (see Table 4). As shown in Table 4, the predicted hydride compounds exhibit superconductivity with a maximum $T_c \approx 100$ K at *ambient pressure*.

The exploration of hydrogen-rich superconductors has evolved through distinct phases. Early experimental efforts focused on binary hydrides such as PdH ($T_c$ = 9 K)[64], $TiH_{0.71}$ ($T_c$ = 4.3 K)[65], $MoH_{1.2}$ ($T_c$ = 0.92 K)[66], $Th_4H_{15}$ ($T_c$ = 8.2 K)[67], $NbH_{x<0.7}$ ($T_c$ = 9.4 K)[68], and $ZrH_3$ ($T_c$ = 11.6 K)[69]. However, progress stagnated for decades, with $T_c$ values constrained to near 10 K. Ternary and multinary hydrides, offering expanded greater compositional and structural diversity than binary counterparts, emerged as promising candidates for achieving elevated $T_c$ at low or ambient pressures[70,71]. Early examples include $HfV_2H$ ($T_c$ = 4.8 K)[72] and $Pd_{0.55}Cu_{0.45}H_{0.7}$ ($T_c$ = 16.6 K)[73], though these studies date back to the 1970s. A paradigm shift occurred with theoretical predictions of high-$T_c$ superconductivity in compressed hydrides such as $H_3S$ ($T_c$ = 203 K at 150 GPa)[74-79] and $LaH_{10}$ ($T_c \approx 250$ K at 170 GPa)[77], which rekindled interest in hydrogen-dominated systems. The original insights are credited to Neil Ashcroft, who first proposed in 1968 that high-temperature superconductivity could theoretically occur in metallic hydrogen[80]. Four decades later, in 2004, he further suggested that metallic hydrogen sublattices might be stabilized at more experimentally accessible pressures within hydrogen-rich compounds[81]. However, the extreme pressures required for these phases limited their practical applicability.

While achieving superconductivity in hydrides at ambient pressure remains challenging, recent computational advances have expanded the search to ternary and multinary hydrides, leveraging their compositional flexibility to stabilize high-$T_c$ states under more accessible conditions. For instance, hole-doped $Mg(BH_4)_2$ has been proposed as an ambient-pressure candidate ($T_c \approx$ 140 K)[82], with $T_c$ reaching 98 K at 0.1 holes per formula unit and increasing to ≈140 K with higher doping. The proposed synthesis pathway involves partial substitution of Mg with Na, providing an energetically favorable approach.

Similarly, machine-learning-assisted high-throughput searches predict certain hydrides, such as $Mg_2XH_6$ (X = Rh, Ir, Pd, Pt), which may exhibit conventional superconductivity with $T_c$ up to 80 K at ambient pressure[83]. These compounds are thermodynamically stable and share structural similarities with experimentally synthesized $Mg_2RuH_6$ (a semiconducting) [84,85]. Upon electron doping (one electron per formula unit in $Mg_2IrH_6$ or two per formula unit in $Mg_2PtH_6$), a superconducting state emerges.

Further ML-assisted studies on hydride superconductors identified ~50 systems with $T_c$ exceeding 20 K, some reaching above 70 K (up to 86 K) at ambient pressure[86]. These systems often combine alkali/alkali-earth elements with noble metals, aligning with the composition of $SM_2TMH_6$ (simple metal-transition metal)[86]. Additionally, $SrNH_4B_6C_6$, a boron-carbon clathrate doped with ammonium hydride units, is computationally predicted to achieve $T_c \approx$ 85–115 K at ambient pressure, leveraging hydrogen's light mass to enhance phonon-mediated pairing[87]. Another proposed metastable compound, cubic $Mg_2IrH_6$, is theorized to reach $T_c \approx$ 160 K, though its synthesis requires high-pressure precursors[88]. These predictions yet underscore the untapped potential of engineered hydride architectures.

Here, our ML-assisted high-throughput studies predict over 40 *ambient*-pressure hydrides with maximum $T_c$ value up to 100 K, including 21 systems exceeding 64 K. As research advances toward ambient high-$T_c$ superconductors, these findings establish a conceptual and methodological framework for guiding targeted discovery and advancing high-$T_c$ superconductivity at ambient pressure.

Our structural analyses, however, reveal a challenge. Preliminary investigations indicate molecular configurations, such as ring-shaped or cluster formations, in select hydrogen-containing compounds (e.g., Figure S9, for instance). These motifs promote localized electronic states and suppress Cooper pair formation, key requirements for superconductivity, suggesting limited superconducting potential in such architectures. To support this assessment, we performed DFT calculations, which revealed small band gaps in these structures, further supporting their poor superconducting viability. This duality highlights the necessity of conducting further experimental and computational studies in future designs. A more detailed quantitative study will follow in future work.

While no hydride has yet demonstrated unambiguous ambient-pressure superconductivity with high $T_c$, and experimental realization of our proposed materials remains pending, this work, along with previous theoretical advances and emerging synthesis strategies, suggests this milestone may be within reach.

Our AI-driven analysis identifies Pu and H as pivotal elements in two distinct superconductor families. While H is empirically recognized in some superconducting systems, its role as a fundamental building block at *ambient* pressure has never been theoretically or computationally established. Our work fills this gap, offering an AI-supported framework that expands the understanding of H's contributions to superconductivity.

Critically, we propose a groundbreaking hypothesis: H may serve as a universal building block for superconductivity at ambient pressure, enabling $T_c$ near 100 K, a threshold previously unattained without extreme pressure. If validated, this insight could redefine the search for high-temperature superconductors, accelerating progress toward room-temperature applications and transformative technological advances.

### 2.4.C Crystal Structure of Predicted Superconductors

We systematically evaluate the crystal structures and physical properties of selected predicted superconductors from two material families: Pu- and H-based compounds (summarized in Tables 4 and 5) in Supplementary information. These systems were analyzed to assess their structural suitability for superconductivity, focusing on lattice configurations and bonding environments conducive to Cooper pair formation. Detailed crystallographic data are provided in Supplementary Information.

## CONCLUSION

This study demonstrates that high-throughput screening (HTS) combined with artificial intelligence (AI) models can significantly expedite the discovery of governing principles in materials science, particularly for identifying fundamental building blocks, or "materials genes," that dictate the properties of superconducting materials. The identification of these building blocks opens new avenues for the synthesis and optimization of novel superconducting materials. By integrating large-scale databases, such as ICSD, Materials Project and COD, into our HTS pipeline, we successfully identified potential high-temperature superconductors. We predicted *two* distinct families of superconductors, with Pu and H identified as the genes of these families. The maximum $T_c$ for predicted compounds within these families is approximately $T_c$ ~ 100 K. We predicted 24 new binary Pu-based superconductors with $T_c$ > 50 K at ambient pressure, including 16 compounds with $T_c$ > 70 K, in three categories: (i) Pu combined with transition elements such as Y, Zr, Os, Co and Mn. (ii) $Pu_5X_3$ compounds (X=Ir, Ru, Os) which Pu combined with three transition elements and (iii) the $Pu_3X'$ series (X'=Yb, La, Dy, Ce, Th, Nd, Er, Pm, Tm, Ho, Sm and Tb), combined with 11 Lanthanide group elements, with $T_c$ values ranging from 65 K ($Pu_3Tb$) to 83 K ($Pu_3Yb$). Additionally, a compound containing Pu and K is also predicted. Our findings reveal that materials containing the elements Plutonium and Hydrogen are particularly promising candidates, exhibiting high transition temperatures and underscoring their potential role as critical components in the design of new superconductors.

These results illustrate the transformative potential of AI-driven HTS in accelerating the discovery of next-generation superconductors.

The predicted results (the introduced superconducting genes) align closely with existing experimental data, demonstrating strong consistency with established superconducting frameworks. However, the predictions proposed here, while theoretically grounded, represent novel hypotheses that necessitate rigorous experimental validation or complementary theoretical corroboration to confirm their viability.

## METHODS AND DATA COLLECTION

### 1. Neural Network Development Using TensorFlow and Keras: Comparing Sequential and Functional API Approaches.

Recent advancements have yielded software libraries that significantly simplify and accelerate neural network research and application. TensorFlow has emerged as a leading framework, greatly enhancing neural network model development. Keras, a high-level Python API built on TensorFlow, further simplifies neural network design, training, and analysis[8,89-92]. We utilized TensorFlow 2.17.0 and Keras 3.4.1 to develop artificial neural networks, leveraging their robust functionalities for effective model implementation.

Keras offers two primary approaches for neural network construction: the Sequential API and the Functional API. The Sequential API suits simple network architectures with single inputs, allowing straightforward layer-by-layer model building. However, it lacks flexibility for complex topologies involving multiple inputs, outputs, or shared layers. The Functional API, meanwhile, offers greater flexibility and supports intricate architectures with multiple inputs, outputs, and shared layers, providing greater versatility[8,91,92].

We developed various neural network models using both APIs, noting that the Functional API generally yielded lower error rates. Consequently, we focus our reporting on the Functional API, which demonstrated superior accuracy and performance.

### 2. Developing Effective Models for Classifying Superconducting from Non-Superconducting Materials

To identify the "materials genes" or fundamental building blocks responsible for superconductivity, we first focused on finding ML models capable of extrapolating to higher transition temperatures. This step is crucial for efficient HTS of large material databases. Next, we aimed to develop ML models capable of classifying materials as superconducting or non-superconducting. To achieve this, a dataset that includes both superconducting and non-superconducting materials is essential, enabling the model to learn the distinguishing features between these two categories. In the following section, we address this challenge by developing and compiling a comprehensive dataset specifically for non-superconducting materials, complementing the existing superconducting dataset. This combined dataset serves as the foundation for training a reliable and precise classification model.

**2.1 New Dataset for Non-Superconducting Materials**

A significant challenge in developing a classification model is the lack of a comprehensive dataset for non-superconducting materials. In contrast, superconducting materials benefit from a relatively comprehensive SuperCon dataset. To address this challenge, some researchers[28,29,44] have leveraged the findings of Hosono and colleagues[93], who examined over 1,000 materials and found that only 3% exhibited superconducting properties. Based on this observation, these researchers proposed that materials present in large databases like COD or Materials Project, but absent from the Supercon database, can be considered as non-superconducting. Another group of researchers[19,43] employed a band gap threshold, classifying materials with a band gap greater than 0.1 eV, as well as insulators and semiconductors, as non-superconducting.

In this study, we expand on the second approach and introduce a new methodology to compile a dataset of non-superconducting materials. We collected materials with known band gaps from the Materials Project and AFlow databases, merged the data, and removed duplicates, resulting in a final dataset of 130,226 unique materials with band gap information.

By comparing this dataset with the SuperCon database, we identified 1,452 superconducting materials in SuperCon that overlap with our dataset and have defined band gaps. Notably, 1,386 of these (over 95%) exhibited a band gap of exactly zero, emphasizing the prevalence of zero band gaps in superconductors. As shown in Figure 9, only 54 of the 1,452 superconducting materials had a band gap exceeding 0.1 eV. This suggests that materials with a band gap greater than 0.1 eV can be classified as non-superconducting. Based on this criterion, we selected 49,163 materials from 130,226 materials as non-superconducting materials. However, this threshold is inadequate, as some non-superconducting materials possess band gaps below 0.1 eV. The following section provides evidence supporting this assertion.

In the SuperCon dataset, the $T_c$ for approximately 4,000 materials is not reported. Following several studies[24,28,29,36], we assigned a $T_c$ of zero (0 K) to these materials, indicating that they are non-superconducting. To validate this classification, we randomly selected and examined references for 30 of the 4,033 materials with unreported $T_c$. Our examination confirmed that none exhibited a superconducting phase. Among the 4,033 non-superconducting materials in the SuperCon dataset, 1,206 are also found in the combined database of Materials Project and AFlow (comprising 130,226 materials) with determined band gaps. Figure 10 illustrates the band gap distribution for these 1,206 materials, showing that the majority exhibit a band gap of zero or less than 0.1 eV.

Consequently, it is reasonable to compile a relatively comprehensive dataset of non-superconducting materials, including both those with band gaps greater than 0.1 eV and those with band gaps of zero or below 0.1 eV. Building on this approach, we added these 4,033 materials to the 49,163 collected from the Materials Project and AFlow, resulting in a total of 53,196 non-superconducting materials. This dataset, called DataG Non-Sc, is now accessible to the broader research community for further study and exploration.

## 3. Data Collection and Pre-Processing

To construct and develop effective classification models, we utilized two distinct datasets: the DataG dataset[33], derived from the largest dataset of superconducting materials, the SuperCon dataset, underwent various stages of data pre-processing[33], containing 13,022 superconducting materials, and the DataG Non-Sc dataset, comprising 53,196 non-superconducting materials. For model training and evaluation, we assigned a label of 1 to superconducting materials and a label of 0 to non-superconducting materials.

We employed both Magpie[94] and Jabir[33] atomic descriptors to generate the feature space for materials. Magpie provides 132 features for each material, whereas Jabir generates 322 features per material. This integration resulted in a total of 454 features for the neural networks during the intermediate stage of model development. Using the Soraya[33] Python package, we identified 30 significant features for feature selection.

## 4. Machine Learning Models and Training

We employed various ML models, including ten traditional ML models: CatBoost, XGBoost, SVM, Bagging, AdaBoost, Decision Tree, Gradient Boosting, Random Forest, KNN and ElasticNet, and nine neural network models: the Functional and Sequential Convolutional Neural Networks (CNN), Long Short-Term Memory (LSTM), the Functional and Sequential Multi-Layer Perceptron (MLP), GRU, Autoencoder and RNN, for classification and prediction tasks.

The evaluation process involved splitting the dataset into training (90%) and testing (10%) sets. To evaluate the performance and extrapolation capabilities of the models, we allocated 5% of the superconducting materials with the highest $T_c$ to the test dataset and used the remaining 95% as training data.

## 5. Performance Metrics

Several performance metrics were utilized, including precision, accuracy, recall, the F1 score, RMSE, and $R^2$. The definitions of these performance metrics are provided by the following formulas[24,29] :

$$Accuracy = \frac{TP+TN}{TP+TN+FP+FN}\ ,\ Precision = \frac{TP}{TP+FP}\ ,\ Recall = \frac{TP}{TP+FN}\ ,$$

$$F1\ Score = 2 * \frac{Precision * Recall}{Precision + Recall}$$

where: TP = True Positives, TN = True Negatives, FP = False Positives, FN = False Negative. Using the confusion matrix, we can evaluate the performance of classification models by calculating metrics such as error rate and accuracy. The confusion matrix contains four key parameters: True Positives (TP), True Negatives (TN), False Positives (FP), and False Negatives (FN).

- True Positive (TP): The number of correctly predicted superconducting materials.
- True Negative (TN): The number of correctly predicted non-superconducting materials.
- False Positive (FP): The number of non-superconducting materials incorrectly classified as superconducting.
- False Negative (FN): The number of superconducting materials incorrectly classified as non-superconducting.

To ensure reliability, we conducted 10-fold cross-validation, dividing the dataset into ten subsets. This method allows the models to be trained and tested multiple times, ensuring robustness. The evaluation metrics were averaged across all folds, with the results presented in Table 6 (top and bottom), based on features derived from Jabir and Magpie descriptors.

In the preceding 10-fold cross-validation step, the optimal set of hyperparameters was identified by selecting the combination that yielded the highest average performance across all folds. The final assessment involved dividing the dataset into a training set (90%) and a testing set (10%).

The ROC (Receiver Operating Characteristic) curve is used to quantify the model's ability to distinguish between superconducting and non-superconducting materials, plotting the True Positive Rate (TPR) against the False Positive Rate (FPR) at various threshold settings. The AUC (Area Under the Curve) quantifies the model's ability to distinguish between classes, where a higher AUC (closer to 1) indicates better classification performance.

## 6. Ensemble Learning

Ensemble learning, inspired by the principle that group decisions often outperform individual ones, is a ML technique that combines multiple models to enhance prediction accuracy and robustness. By aggregating outputs from diverse algorithms, ensemble methods leverage their strengths to improve classification or regression outcomes, reducing errors and increasing reliability. This approach is particularly effective in complex tasks, such as distinguishing superconducting from non-superconducting materials, where multiple models provide diverse perspectives for more informed predictions[95-97].

To leverage the advantages of ensemble learning, we selected two traditional ML models (XGBoost and CatBoost) and two neural network models (CNN and MLP). A material was classified as superconducting only if at least two of the four models independently identified it as such. This stringent criterion minimized misclassification, enhancing the reliability of our results.

## 7. High Throughput Screening (HTS) Pipeline

Our HTS pipeline analyzed data from extensive material databases, such as the ICSD, Materials Project and COD using AI-based models. We integrated the developed ML models into the pipeline, allowing the identification of promising high-temperature superconducting materials.

This approach facilitated the discovery of fundamental building blocks for superconducting materials, such as Plutonium, which exhibited high $T_c$ values.

The innovative approach introduced in this article, discovering building blocks of material properties through HTS using AI-based models, has the potential to inspire researchers in the field of materials science. This methodology may facilitate the discovery of new materials with unique properties.

**Data availability**

To address the scarcity of data on non-superconducting materials, we compiled a publicly accessible dataset (DataG Non-Sc) comprising 53,196 materials. This dataset encompasses materials with band gaps both above and below 0.1 eV, curated using the methodologies described in this study. DataG Non-Sc is openly available to facilitate further research and can be accessed via the GitHub repository: https://github.com/Hassan-Gashmard/DataG-Non-Sc

**Acknowledgment**

This work was supported by the Iran National Science Foundation (INSF). We gratefully acknowledge their financial support. The authors also thank Dr. T. Morshedloo for her helpful discussions.

# Tables

**Table 1**: **Evaluation Metrics for Traditional ML Models Trained using Jabir-Generated Features**

**Table 2**: **Average Evaluation Metrics from 10-Fold Cross-Validation of CNN and MLP Models for both training and testing data.** Results were derived using features generated by the Jabir package (*top*) and Magpie descriptors (*bottom*). Metrics are reported for both training and testing data.

**Table 3**: **Evaluation metrics for CNN and MLP models trained using feature space generated by the Jabir package.** The dataset was split into 90% training and 10% testing subsets for final model evaluation.

**Table 4**. **Predicted High-Temperature Superconductors using Neural Network Models at Ambient Pressure.** Data from the Inorganic Crystallographic Structure Database (ICSD), Materials Project (MP) and Crystallography Open Database (COD) were utilized. A dual-model architecture incorporating Convolutional Neural Networks (CNNs) and Long Short-Term Memory (LSTM) networks were developed. CNNs demonstrated superior performance in extrapolating higher transition temperatures ($T_c$). The Pu and H elements, identified as the predicted Gens in the certain predicted superconductor families. Also, the predicted Pu-based compounds were compared with experimental data. *Note:* the selected predicted compounds with nearly high $T_c$ are included on the table and others have not been shown.

**Table 5**. **The crystal structure parameters and some of the physical properties for a few predicted High-Temperature Superconductors in this study.** The table shows a few compounds of two certain predicted family of superconductors, Pu- and H-based compound. St.F.: Standard formula, SG: Space group, BCT: body-centered tetragonal.

**Table 6**: **Evaluation Metrics for Training and Testing Data using 10-fold Cross-Validation across Traditional ML Algorithms.** The values represent the average performance across all folds, highlighting the robustness of each classification model. Results demonstrated using the features generated by Jabir (*top*) and Magpie descriptors (*bottom*).

## Figures

**Figure. 1 A schematic illustrating the prediction of high-$T_c$ superconductors using HTS techniques and advanced ML models.** Functional CNNs are applied to extrapolate potential high-$T_c$ compounds. Leveraging datasets like ICSD, the Materials Project, and COD, our HTS pipeline classifies superconductors, while CNN and LSTM models predict $T_c$ and elemental composition. Two predicted families, including Pu and H, have been identified.

**Figure 2. a) Root Mean Squared Error (RMSE) and b) Mean Absolute Error (MAE) for traditional machine learning models evaluating extrapolation of transition temperatures for superconducting materials.**

**Figure 3. Number of compounds and the percentage of successful transition temperature extrapolation within the test DataG dataset using traditional ML algorithms (651 compounds with the highest $T_c$ in the dataset).**

**Figure 4. a) Root Mean Squared Error (RMSE) and b) Mean Absolute Error (MAE) for performing various neural network models, employing both sequential and functional API approaches, evaluating extrapolation of transition temperatures for superconducting materials.**

**Figure 5. Analysis of the number of compounds and the percentage of successful transition temperature extrapolation within the test data using artificial neural network models (651 compounds with the highest $T_c$ in the dataset).**

**Figure 6. Confusion matrices for the CatBoost model using features generated from a) Jabir and b) Magpie descriptors.**

**Figure 7. Confusion matrices for the Convolutional Neural Network (CNN) model evaluated using features derived from the a) Jabir and b) Magpie descriptors.**

**Figure 8. Comparative analysis of superconducting transition temperature predictions using a) $R^2$ evaluation metric and b) Root Mean Square Error (RMSE) across various traditional ML and neural network models.**

**Figure 9. Comparative analysis of the SuperCon dataset and 130,226 collected materials from Materials Project and AFlow databases, identifying 1,452 overlapping superconducting materials with defined band gaps.** Notably, over 95% (1,386) of these superconductors exhibited zero band gap, while only 54 materials had a band gap exceeding 0.1 eV.

**Figure 10. Band gap distribution for 1,206 non-superconducting materials from the SuperCon dataset, overlapping with the combined Materials Project and AFlow database of 130,226 materials.**

**Table 1. Evaluation Metrics for the Traditional ML Models Trained using Jabir-Generated Features**

| Model | Accuracy (train) | Accuracy (test) | Precision (train) | Precision (test) | Recall (train) | Recall (test) | F1-Score (train) | F1-Score (test) |
|---|---|---|---|---|---|---|---|---|
| CatBoost | 1.00 | 0.9998 | 1.00 | 1.00 | 1.00 | 0.9992 | 1.00 | 0.9996 |
| XgBoost | 1.00 | 0.9995 | 1.00 | 0.9992 | 1.00 | 0.9984 | 1.00 | 0.9988 |
| Random Forest | 1.00 | 0.9992 | 1.00 | 0.9976 | 1.00 | 0.9984 | 1.00 | 0.9980 |
| Decision Tree | 0.9997 | 0.9992 | 0.9995 | 0.9984 | 0.9991 | 0.9976 | 0.9993 | 0.9980 |
| KNN | 0.9685 | 0.9450 | 0.9432 | 0.8795 | 0.8937 | 0.8396 | 0.9178 | 0.8591 |
| Logistic | 0.9052 | 0.8998 | 0.8790 | 0.8741 | 0.6001 | 0.5827 | 0.7132 | 0.6993 |

**Table 2. Average Evaluation Metrics from 10-Fold Cross-Validation of CNN and MLP Models for both training and testing data.** Results were derived using features generated by the Jabir package (*top*) and Magpie descriptors (*bottom*). Metrics are reported for both training and testing data.

| Model | Accuracy (train) | Accuracy (test) | Precision (train) | Precision (test) | Recall (train) | Recall (test) | F1-Score (train) | F1-Score (test) |
|---|---|---|---|---|---|---|---|---|
| CNN | 0.9957 | 0.9956 | 0.9854 | 0.9852 | 0.9931 | 0.9926 | 0.9892 | 0.9889 |
| MLP | 0.9947 | 0.9944 | 0.9818 | 0.9817 | 0.9913 | 0.9900 | 0.9865 | 0.9857 |

| Model | Accuracy (train) | Accuracy (test) | Precision (train) | Precision (test) | Recall (train) | Recall (test) | F1-Score (train) | F1-Score (test) |
|---|---|---|---|---|---|---|---|---|
| CNN | 0.9287 | 0.9276 | 0.7533 | 0.7512 | 0.9481 | 0.9452 | 0.8395 | 0.8370 |
| MLP | 0.9311 | 0.9287 | 0.7579 | 0.7528 | 0.9549 | 0.9496 | 0.8450 | 0.8397 |

**Table 3. Evaluation metrics for CNN and MLP models trained using feature space generated by the Jabir package.** The dataset was split into 90% training and 10% testing subsets for final model evaluation.

| Model | Accuracy (train) | Accuracy (test) | Precision (train) | Precision (test) | Recall (train) | Recall (test) | F1-Score (train) | F1-Score (test) |
|---|---|---|---|---|---|---|---|---|
| CNN | 0.9994 | 0.9992 | 0.9980 | 0.9976 | 0.9992 | 0.9984 | 0.9986 | 0.9980 |
| MLP | 0.9997 | 0.9996 | 0.9989 | 0.9992 | 0.9995 | 0.9992 | 0.9992 | 0.9992 |

**Table 4. Predicted HTSC compound using Neural Network Models at Ambient Pressure.**

| Predicted New Sc compound | $T_c$ (K) predicted by CNN | $T_c$ (K) predicted by LSTM | Dataset | Compound | $T_c$ (K) measured | ref |
|---|---|---|---|---|---|---|
| $Pu_{28}Zr$ | 100.5 | 95.9 | ICSD | $PuCoGa_5$ | 18.5 | 54 |
| $Pu_{19}Os$ | 99.1 | 90.7 | ICSD | $PuCo_{0.9}Ni_{0.1}Ga_5$ | 16.6 | 98 |
| $Pu_2Co_{17}$ | 88.6 | 90.0 | ICSD | $PuCo_{0.9}Fe_{0.1}Ga_5$ | 13.5 | 98 |
| $Pu_3Yb$ | 82.8 | 72.0 | MP | $PuCo_{0.8}Fe_{0.2}Ga_5$ | 10 | 98 |
| $Pu_3Y$ | 79.1 | 69.3 | MP | $PuRhGa_5$ | 8.5 | 60 |
| $Pu_3La$ | 76.9 | 75.0 | MP | $PuCo_{0.1}Rh_{0.9}Ga_5$ | 10.2 | 98 |
| $Pu_3Dy$ | 75.4 | 69.6 | MP | $PuCo_{0.5}Rh_{0.5}Ga_5$ | 15.5 | 98 |
| $Pu_3Ce$ | 75.1 | 69.5 | MP | | | |
| $Pu_3Th$ | 72.6 | 69.4 | MP | | | |
| $Pu_3Zr$ | 72.5 | 76.0 | MP | | | |
| $Pu_3Nd$ | 71.5 | 73.6 | MP | | | |
| $Pu_3Er$ | 71.6 | 68.8 | MP | | | |
| $Pu_3Pm$ | 71.5 | 71.5 | MP | | | |
| $Pu_3Tm$ | 71.8 | 73.6 | MP | | | |
| $Pu_3Ho$ | 70.0 | 65.0 | MP | | | |
| $Pu_3Sm$ | 69.7 | 62.5 | MP | | | |
| $Pu_3Tb$ | 65.0 | 58.0 | COD | | | |
| $Pu_5Ir_3$ | 55.6 | 62.0 | MP | | | |
| $Pu_5Ru_3$ | 53.7 | 49.8 | ICSD | | | |
| $Pu_5Os_3$ | 53.6 | 61.8 | ICSD | | | |
| $PuK$ | 52.0 | 42.5 | COD | | | |
| $Pu_4Mn$ | 46.5 | 43.0 | COD | | | |
| $H_{1.7}C_{10}Al_{2.7}FO_{11.2}P_{2.7}$ | 95.8 | 171.1 | ICSD | | | |
| $H_6C_{10}Al_{1.1}NaO_{13}Si_{3.9}$ | 81.4 | 146.8 | ICSD | | | |
| $NaMo_{368}H_{1410}(S_{16}O_{643})_3$ | 57.2 | 71.0 | COD | | | |
| $Zn_3H_{29}C_{42}N_3O_{22}$ | 59.9 | 118 | COD | | | |
| $H_{19}C_{1.5}KO_{16.6}U$ | 49.1 | 90.9 | ICSD | | | |
| $H_{3.7}O_{6.8}U$ | 33.3 | 58.8 | ICSD | | | |
| $H_{3.9}MoO_{4.9}$ | 17.2 | 31.8 | ICSD | | | |
| $H_{49}C_{43}$ | - | 74.6 | COD | | | |
| $H_{50}C_{43}$ | - | 73.8 | COD | | | |
| $H_{49}C_{44}$ | - | 73.5 | COD | | | |
| $H_{0.008}Nb_{0.81}W_{0.008}Zr_{0.182}$ | 9.7 | 8.0 | ICSD | | | |

| | | | | |
|---|---|---|---|---|
| **$HNb_3Sn$** | **9.5** | **6.2** | ICSD | |
| **$H_{0.25}C_{0.7}Nb$** | **8.5** | **9.7** | ICSD | |
| **$H_{0.25}Re$** | **5.4** | **4.8** | ICSD | |
| **$H_{1.4}Mo_2Zr$** | **5.1** | **6.6** | ICSD | |

**Table 5. The crystal structure parameters and some of the physical properties for a few predicted High-Temperature Superconductors in this study.** The table shows a few compounds of two certain predicted family of superconductors, Pu- and H-based compound. St.F.: Standard formula, SG: Space group, BCT: body-centered tetragonal.

| Compound | St.F. | structure | SG | *a* (Å) | *b* (Å) | *c* (Å) | V (Å³) | ref |
|---|---|---|---|---|---|---|---|---|
| $Pu_{28}Zr$ | | BCT | $I4_1/a$ | 18.1899 | 18.1899 | 7.8576 | 2599.86 | 99 |
| $Pu_{19}Os$ | | Orthorhombic | *Cmca(64)* | 5.345 | 14.884 | 10.898 | 866.99 | 100 |
| $Pu_2Co_{17}$ | | Hexagonal | $P6_3/mmc$ | 8.29 (α: 90°) | 8.29 (β:90°) | 8.08 (γ:120°) | 480.91 | 101 |
| $Pu_3Yb$ | | Tetragonal<br>Hexagonal | $I4/mmm$<br>$P6_3/mmc$ | 4.66<br>6.84 | 4.66<br>6.84 | 9.46<br>5.66 | 205.71<br>229.25 | 102 |
| $Pu_3Y$ | | Hexagonal | $P6_3/mmc$ | 6.86 | 6.86 | 5.58 | 227.90 | 102 |
| $Pu_3La$ | | Tetragonal | $I4/mmm$ | 4.91 | 4.91 | 9.86 | 237.53 | 102 |
| $Pu_3Dy$ | | Tetragonal | $I4/mmm$ | 4.83 | 4.83 | 9.68 | 226.03 | 102 |
| $Pu_3Th$ | | Tetragonal<br>Hexagonal | $I4/mmm$<br>$P6_3/mmc$ | 6.90<br>4.69 | 6.90<br>4.69 | 5.44<br>9.46 | 224.01<br>207.90 | 102 |
| $Pu_3Zr$ | | Tetragonal | $I4/mmm$ | 4.60 | 4.60 | 9.00 | 190.03 | 102 |
| $Pu_3Nd$ | | Tetragonal | $I4/mmm$ | 4.86 | 4.86 | 9.89 | 234.16 | 102 |
| $Pu_3Er$ | | Hexagonal | $P6_3/mmc$ | 6.84 | 6.84 | 5.50 | 223.09 | 102 |
| $Pu_3Pm$ | | Tetragonal | $I4/mmm$ | 4.84 | 4.84 | 9.90 | 232.25 | 102 |
| $Pu_3Tm$ | | Tetragonal<br>Hexagonal | $I4/mmm$<br>$P6_3/mmc$ | 4.81 | 4.81 | 9.56 | 221.26 | 102 |
| $Pu_3Ho$ | | Hexagonal | $P6_3/mmc$ | 6.91 | 6.91 | 5.57 | 229.99 | 102 |
| $Pu_3Sm$ | | Tetragonal | $I4/mmm$ | 4.70 | 4.70 | 9.57 | 211.01 | 102 |
| $Pu_3Tb$ | | Hexagonal | $P6_3/mmc$ | 6.93 | 6.93 | 5.57 | 231.58 | 102 |
| $Pu_5Ru_3$ | | Tetragonal | $I4/mcm$ | 10.82 | 10.82 | 5.67 | 664.05 | 102 |
| $Pu_5Os_3$ | | Tetragonal | $I4/mcm$ | 10.96 | 10.96 | 5.54 | 666.12 | 102 |
| $Pu_5Ir_3$ | | Tetragonal | $I4/mcm$ | 11.15 | 11.15 | 5.58 | 693.50 | 102 |
| PuK | | Triclinic | $P\bar{1}$ | 5.99 (α:85.88°) | 6.36 (β:84.80°) | 13.56 (γ:86.99°) | 512.32 | 102 |
| $Pu_4Mn$ | | Cubic | $Fd\bar{3}m1$ | 9.12 | 9.12 | 9.12 | 758.03 | 102 |
| $H_{1.7}C_{10}Al_{2.7}FO_{11.2}P_{2.7}$ | $C_{2826.8}H_{476}Al_{768}F_{283.8}O_{3168}P_{768}$ | Cubic | $Fm\bar{3}c$ | 51.3636 | 51.3636 | 51.3636 | 135508.4 | 103 |
| $H_6C_{10}Al_{1.1}NaO_{13}Si_{3.9}$ | | Cubic | $Im\bar{3}m$ | 44.924 | 44.924 | 44.924 | 90665.29 | 104 |
| $NaMo_{368}H_{1410}(S_{16}O_{643})_3$ | $H_{1410}Mo_{368}NaO_{1929}S_{48}$ | Tetragonal | $I4mm$ | 43.465 | 43.465 | 69.393 | 131096.48 | 105 |
| $Zn_3H_{29}C_{42}N_3O_{22}$ | | Cubic | $Fm\bar{3}c$ | 63.0890 | 63.0890 | 63.0890 | 251108 | 106 |
| $H_{3.7}O_{6.8}U$ | $H_{944}O_{1752}U_{256}$ | Monoclinic | $P\,1\,2_1/c\,1$ | 38.84 (α: 90°) | 36.52 (β:102.863°) | 41.3 (γ: 90°) | 57111.35 | 107 |
| $H_{49}C_{43}$ | | Monoclinic | $P\,1\,2_1/c\,1$ | 16.8988 ( α:90°) | 18.7150 ( β:96.85°) | 24.6740 ( γ:90°) | 7747.7 | 108 |
| $H_{50}C_{43}$ | | Monoclinic | $P\,1\,2_1/c\,1$ | 31.61 (α: 90°) | 15.28 (β:108.29°) | 14.80 (γ: 90°) | 6793.0 | 109 |
| $H_{49}C_{44}$ | | Triclinic | $P1$ | 7.54 (α:86.39°) | 12.02 (β:81.24°) | 19.02 (γ:86.00°) | 1698.5 | 110 |

**Table 6. Evaluation Metrics for Training and Testing Data using 10-fold Cross-Validation across Traditional ML Algorithms.** The values represent the average performance across all folds, highlighting the robustness of each classification model. Results demonstrated using the features generated by Jabir (*top*) and Magpie descriptors (*bottom*).

| Model | Accuracy (train) | Accuracy (test) | Precision (train) | Precision (test) | Recall (train) | Recall (test) | F1-Score (train) | F1-Score (test) |
|---|---|---|---|---|---|---|---|---|
| CatBoost | 1.00 | 0.9999 | 1.00 | 0.9998 | 1.00 | 0.9997 | 1.00 | 0.9998 |
| XgBoost | 1.00 | 0.9999 | 1.00 | 0.9999 | 1.00 | 0.9995 | 1.00 | 0.9997 |
| Random Forest | 1.00 | 0.9995 | 1.00 | 0.9982 | 1.00 | 0.9991 | 1.00 | 0.9987 |
| Decision Tree | 0.9996 | 0.9983 | 0.9998 | 0.9964 | 0.9982 | 0.9949 | 0.9990 | 0.9957 |
| KNN | 0.9689 | 0.9470 | 0.9445 | 0.8911 | 0.8946 | 0.8323 | 0.9189 | 0.8606 |
| Logistic | 0.9029 | 0.9028 | 0.7482 | 0.7486 | 0.7956 | 0.7951 | 0.7640 | 0.7637 |

| Model | Accuracy (train) | Accuracy (test) | Precision (train) | Precision (test) | Recall (train) | Recall (test) | F1-Score (train) | F1-Score (test) |
|---|---|---|---|---|---|---|---|---|
| CatBoost | 0.9877 | 0.9611 | 0.9517 | 0.8827 | 0.9874 | 0.9255 | 0.9692 | 0.9036 |
| XgBoost | 0.9989 | 0.9633 | 0.9960 | 0.8974 | 0.9987 | 0.9185 | 0.9973 | 0.9078 |
| Random Forest | 0.9678 | 0.9513 | 0.8731 | 0.8357 | 0.9784 | 0.9363 | 0.9228 | 0.8831 |
| Decision Tree | 0.9645 | 0.9432 | 0.8846 | 0.8304 | 0.9424 | 0.8940 | 0.9125 | 0.8610 |
| KNN | 0.9697 | 0.9445 | 0.9062 | 0.8376 | 0.9434 | 0.8908 | 0.9244 | 0.8634 |
| Logistic | 0.9027 | 0.9016 | 0.7534 | 0.7508 | 0.7508 | 0.7479 | 0.7520 | 0.7492 |

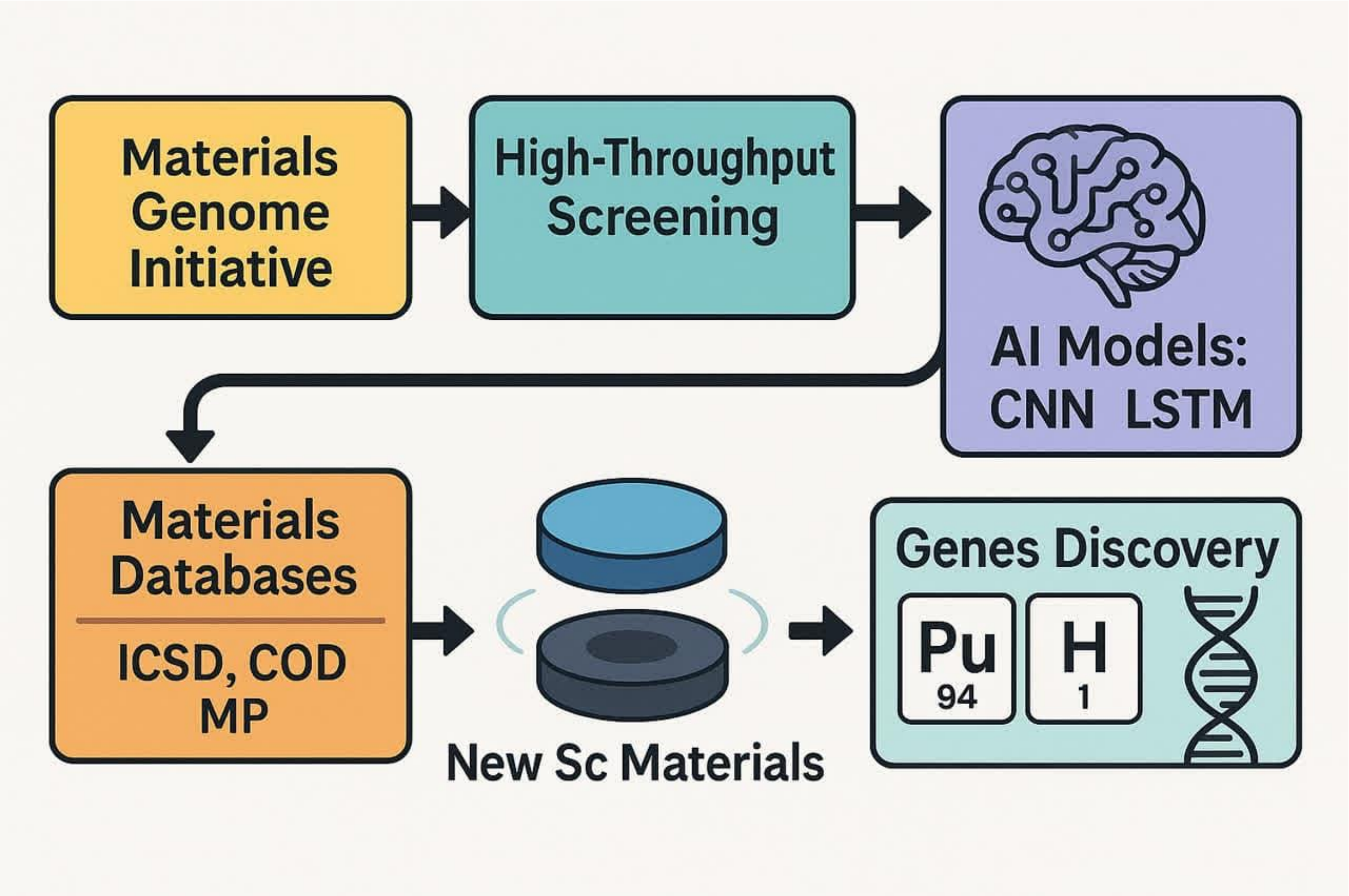

**Figure. 1 A schematic illustrating the prediction of high-$T_c$ superconductors using HTS techniques and advanced ML models.** Functional CNNs are applied to extrapolate potential high-$T_c$ compounds. Leveraging datasets like ICSD, the Materials Project, and COD, our HTS pipeline classifies superconductors, while CNN and LSTM models predict $T_c$ and elemental composition. Two predicted families, including Pu and H, have been identified.

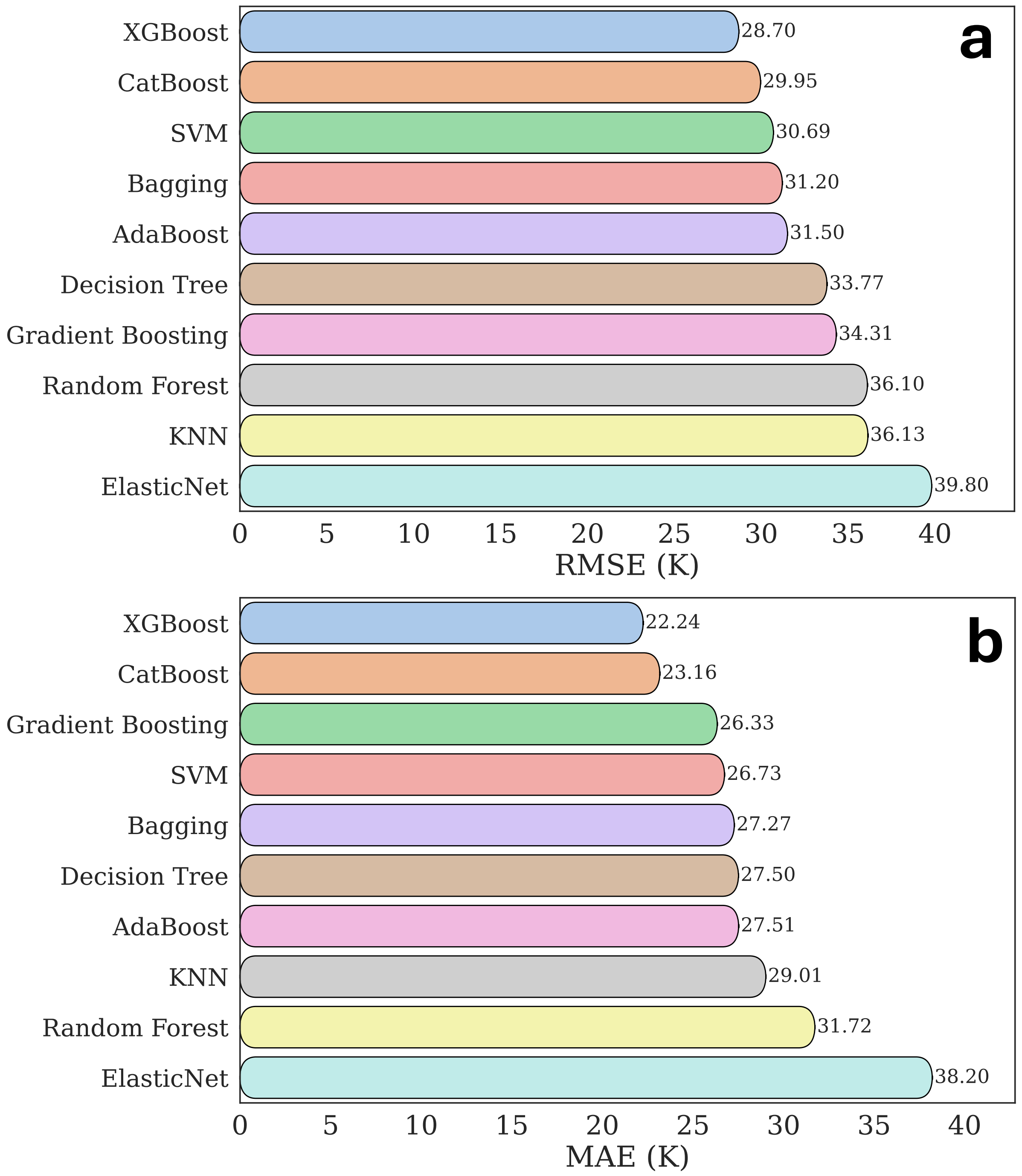

**Figure 2. a) Root Mean Squared Error (RMSE) and b) Mean Absolute Error (MAE) for traditional machine learning models evaluating extrapolation of transition temperatures for superconducting materials.**

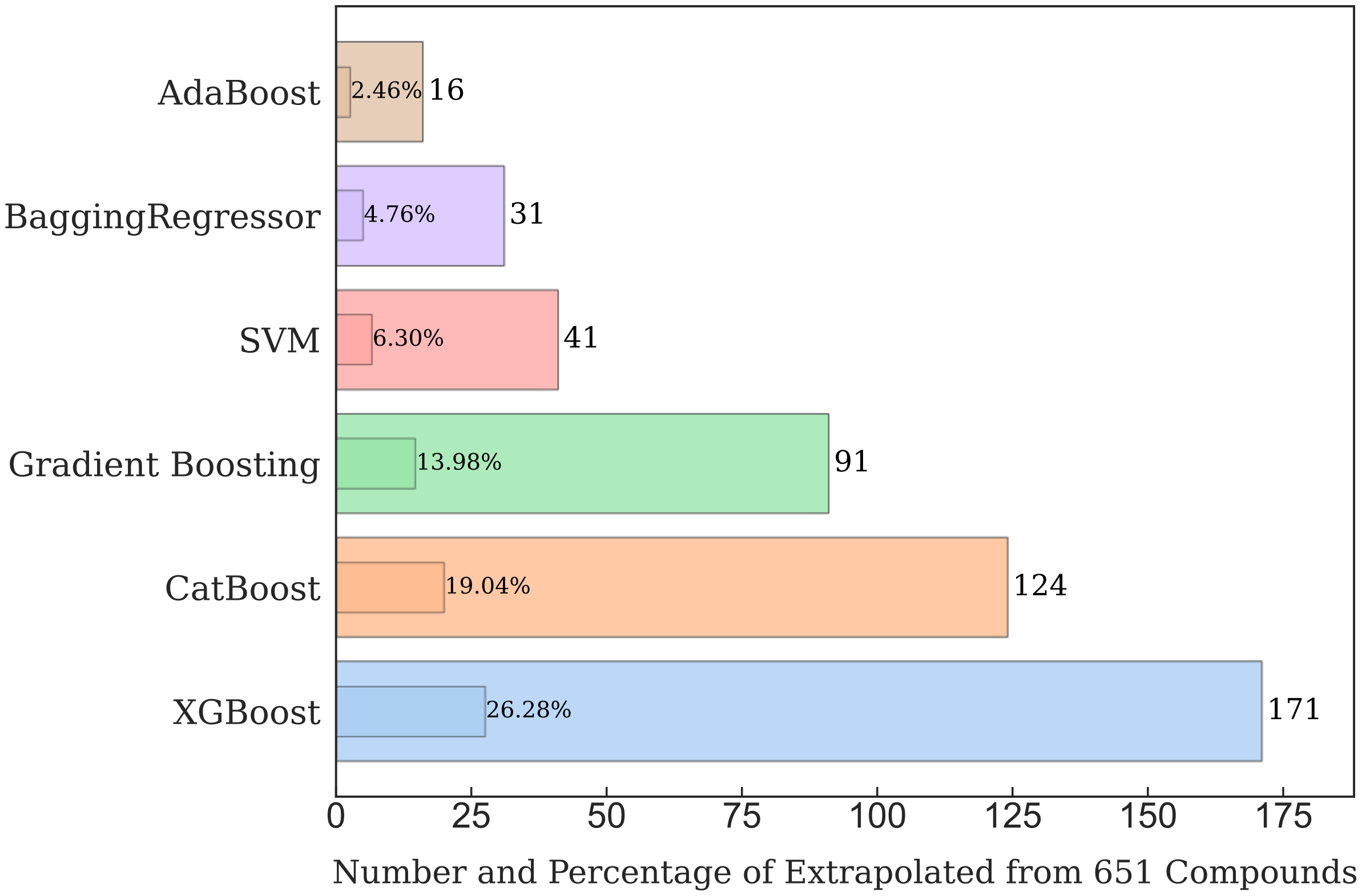

**Figure 3. Number of compounds and the percentage of successful transition temperature extrapolation within the test DataG dataset using traditional ML algorithms (651 compounds with the highest $T_c$ in the dataset).**

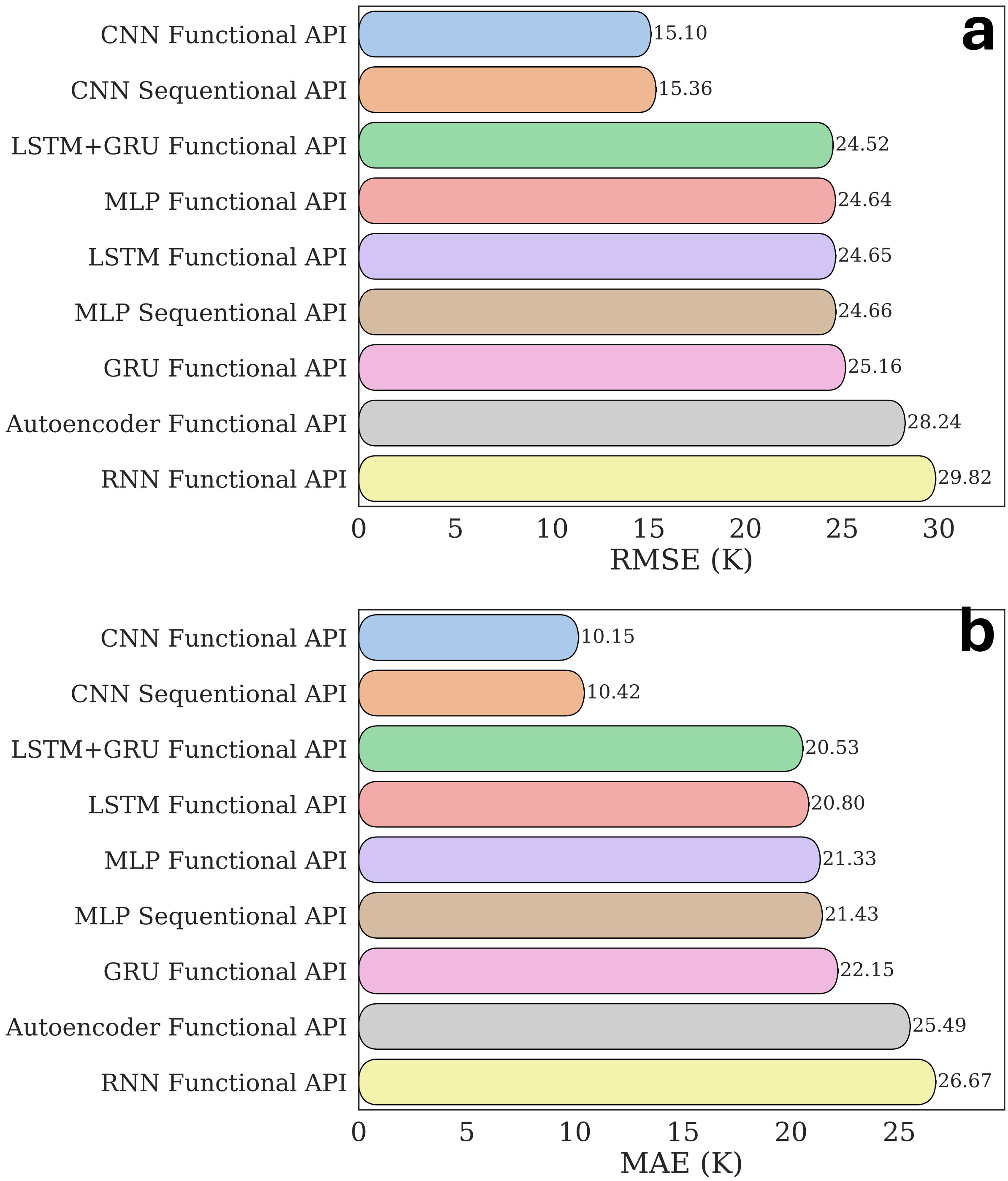

**Figure 4. a) Root Mean Squared Error (RMSE) and b) Mean Absolute Error (MAE) for performing various neural network models, employing both sequential and functional API approaches, evaluating extrapolation of transition temperatures for superconducting materials.**

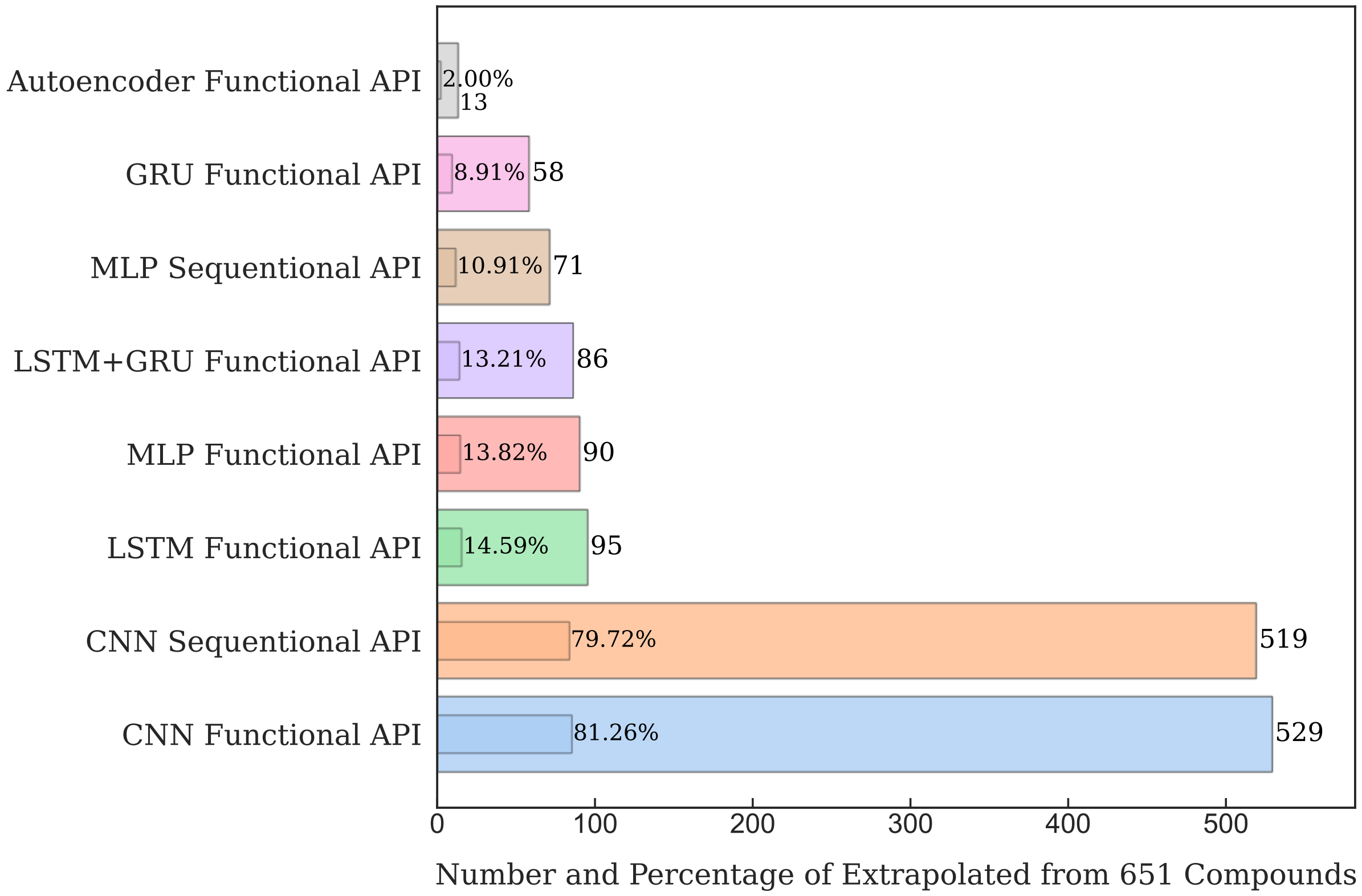

**Figure 5**. **Analysis of the number of compounds and the percentage of successful transition temperature extrapolation within the test data using artificial neural network models (651 compounds with the highest $T_c$ in the dataset).**

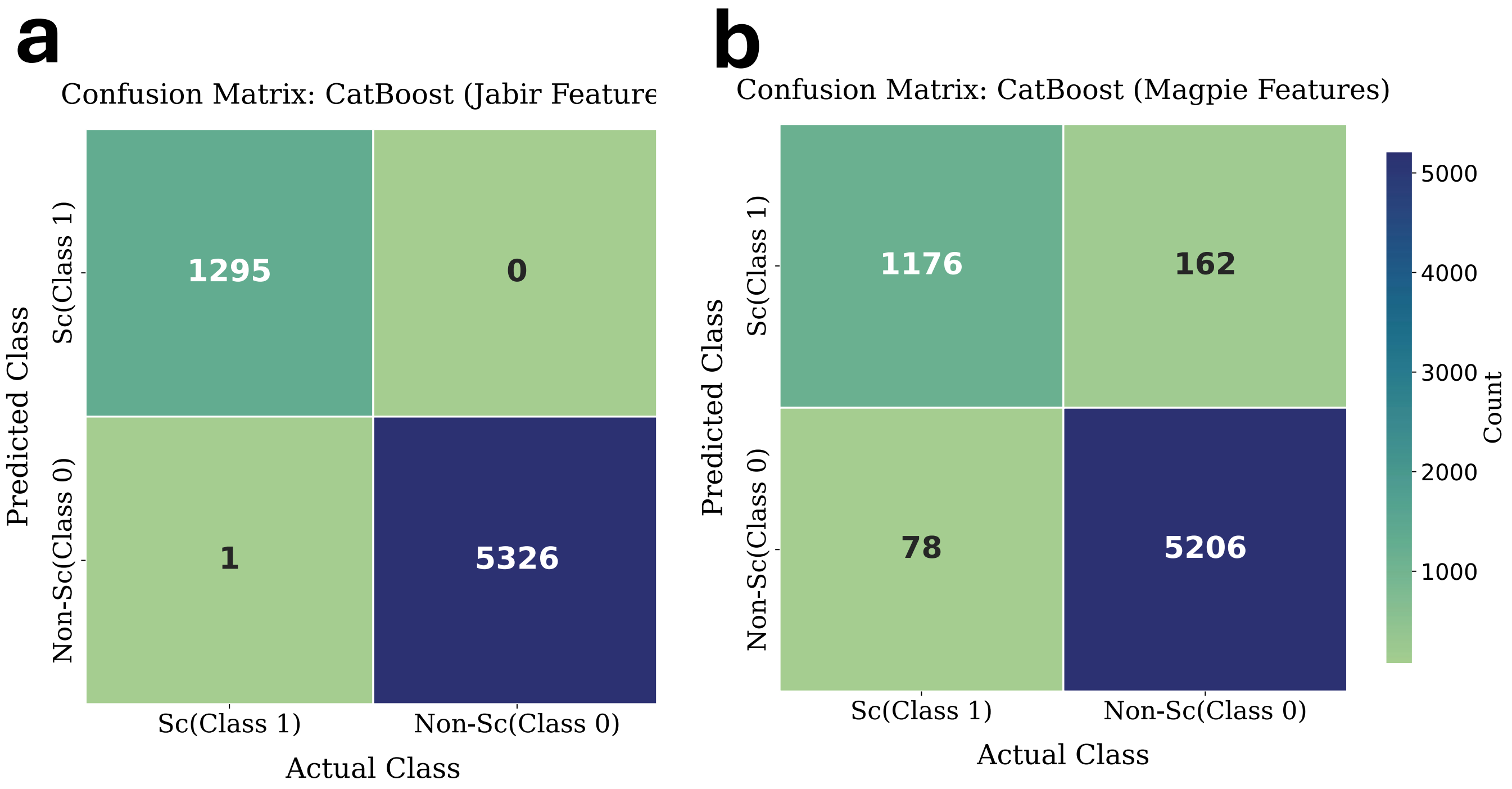

**Figure 6. Confusion matrices for the CatBoost model using features generated from a) Jabir and b) Magpie descriptors.**

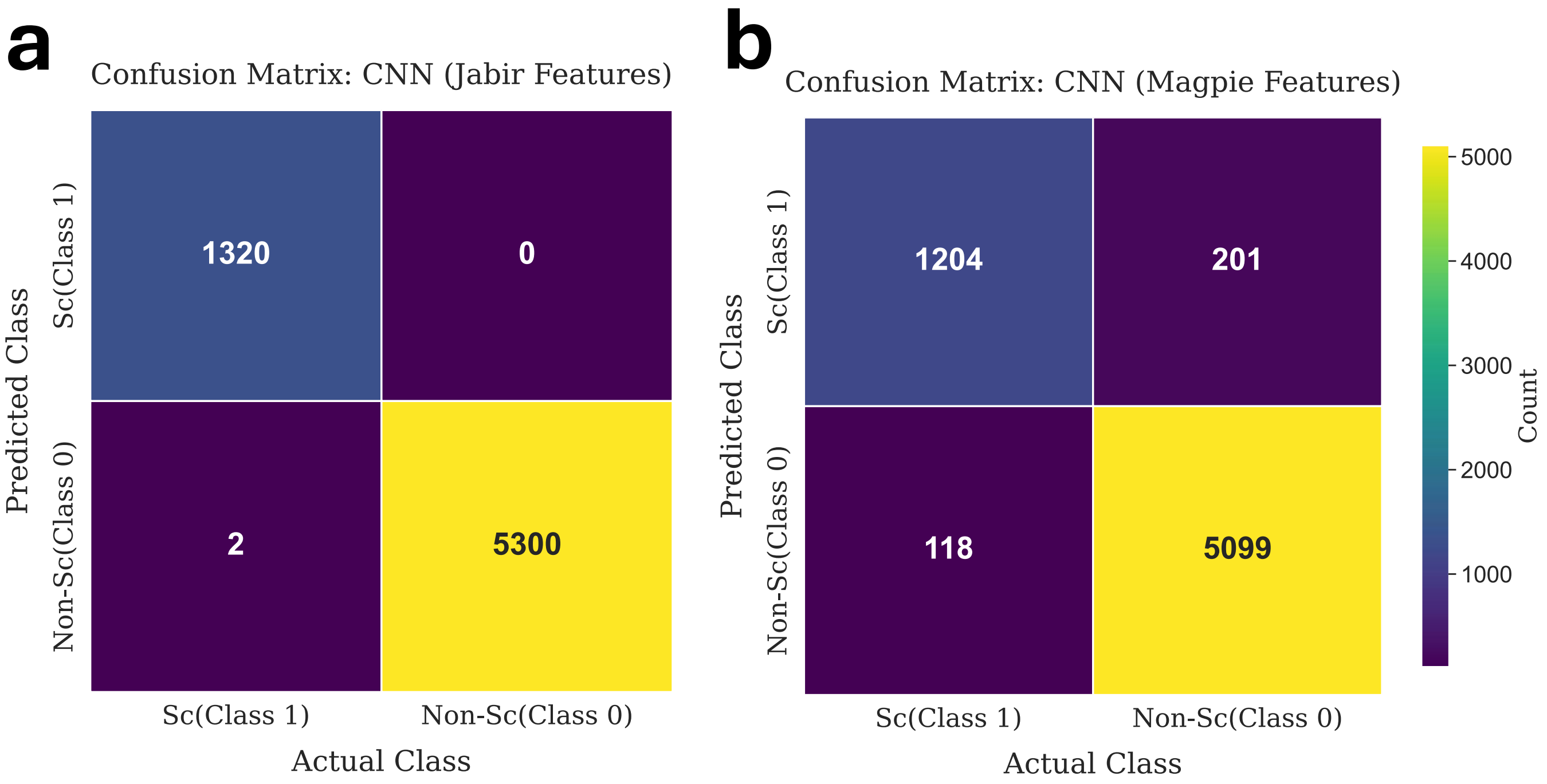

**Figure 7. Confusion matrices for the Convolutional Neural Network (CNN) model evaluated using features derived from the a) Jabir and b) Magpie descriptors.**

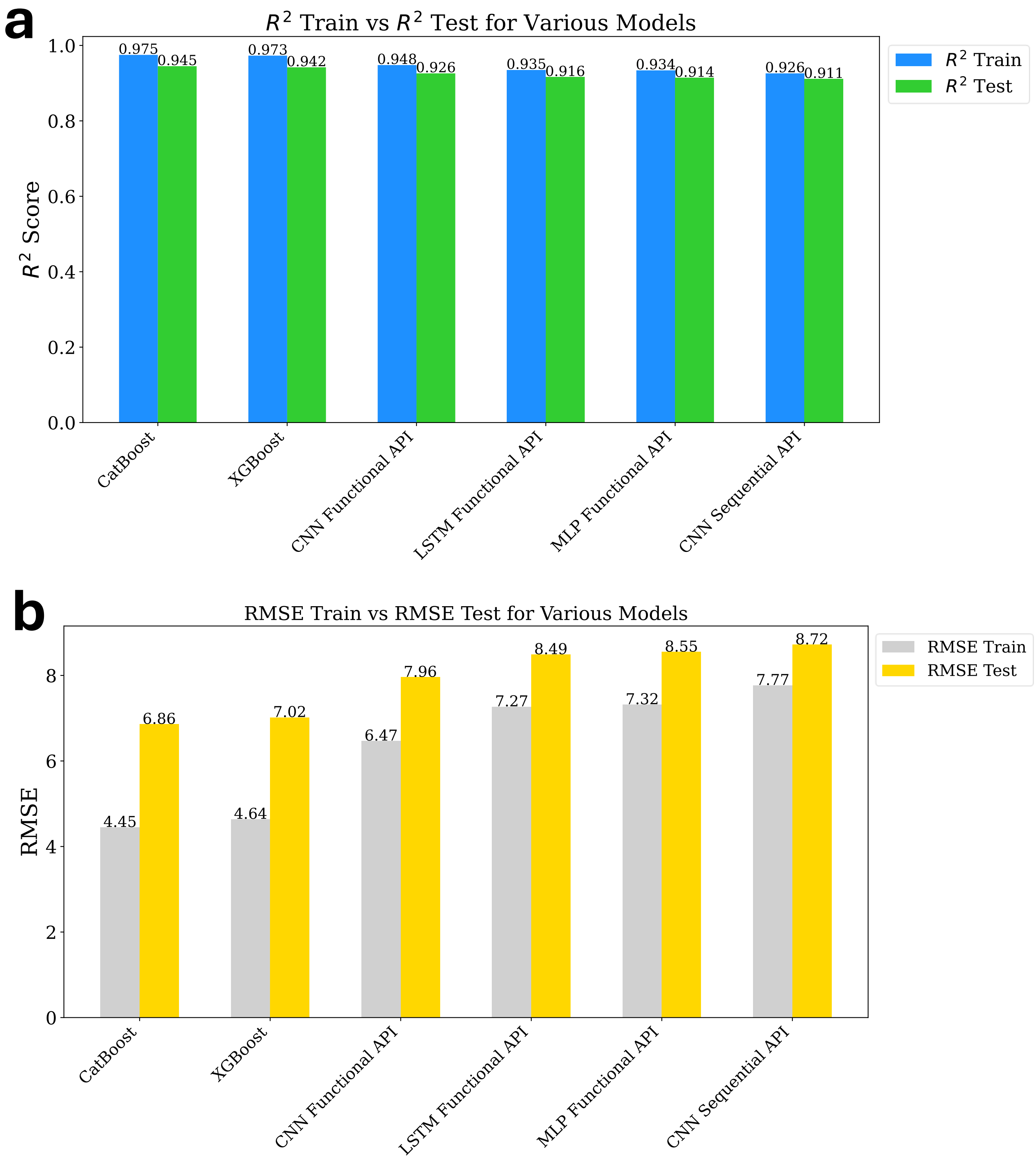

**Figure 8. Comparative analysis of superconducting transition temperature predictions using a) $R^2$ evaluation metric and b) Root Mean Square Error (RMSE) across various traditional ML and neural network models.**

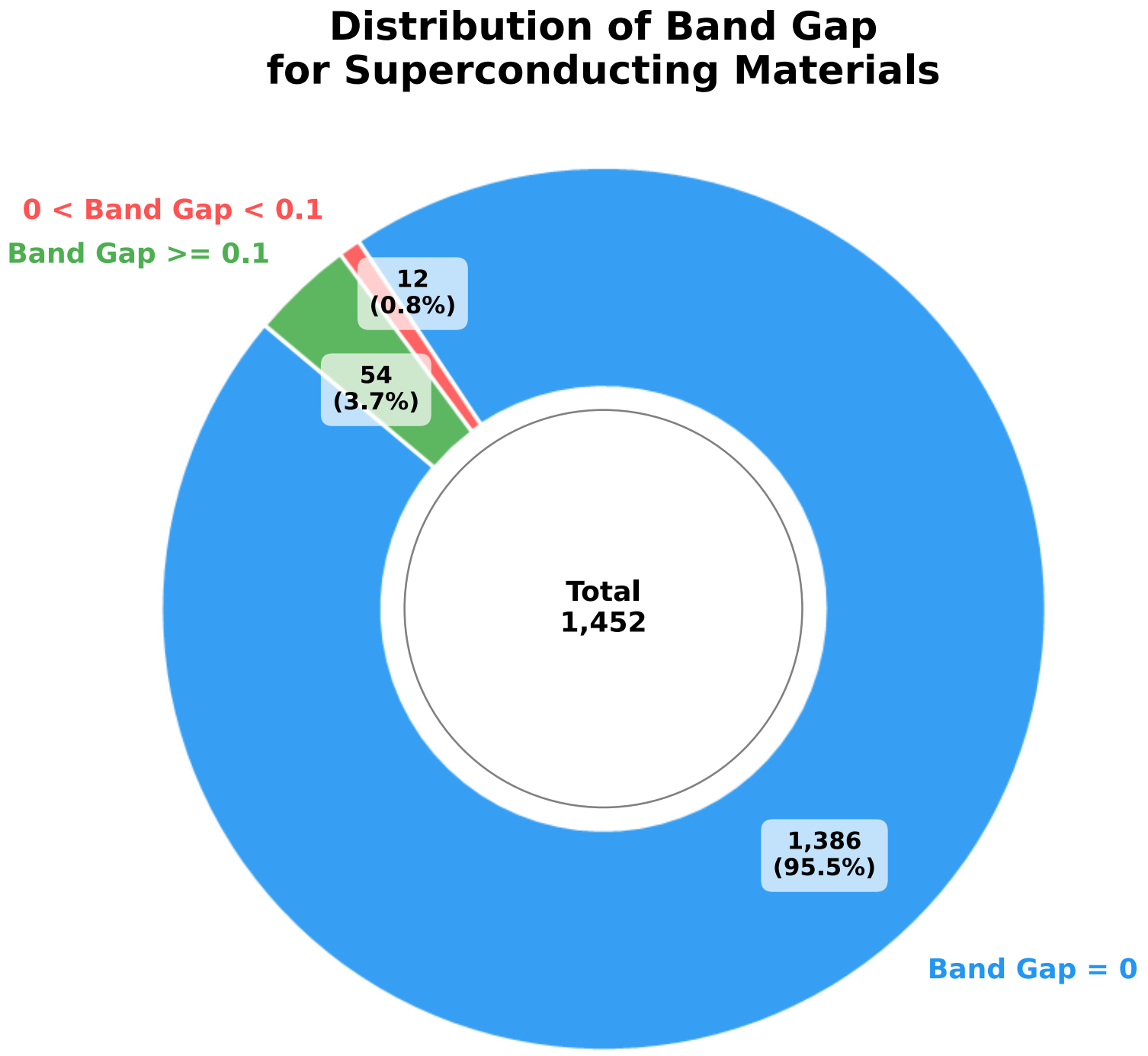

**Figure 9**. **Comparative analysis of the SuperCon dataset and 130,226 collected materials from Materials Project and AFlow databases, identifying 1,452 overlapping superconducting materials with defined band gaps.** Notably, over 95% (1,386) of these superconductors exhibited zero band gap, while only 54 materials had a band gap exceeding 0.1 eV.

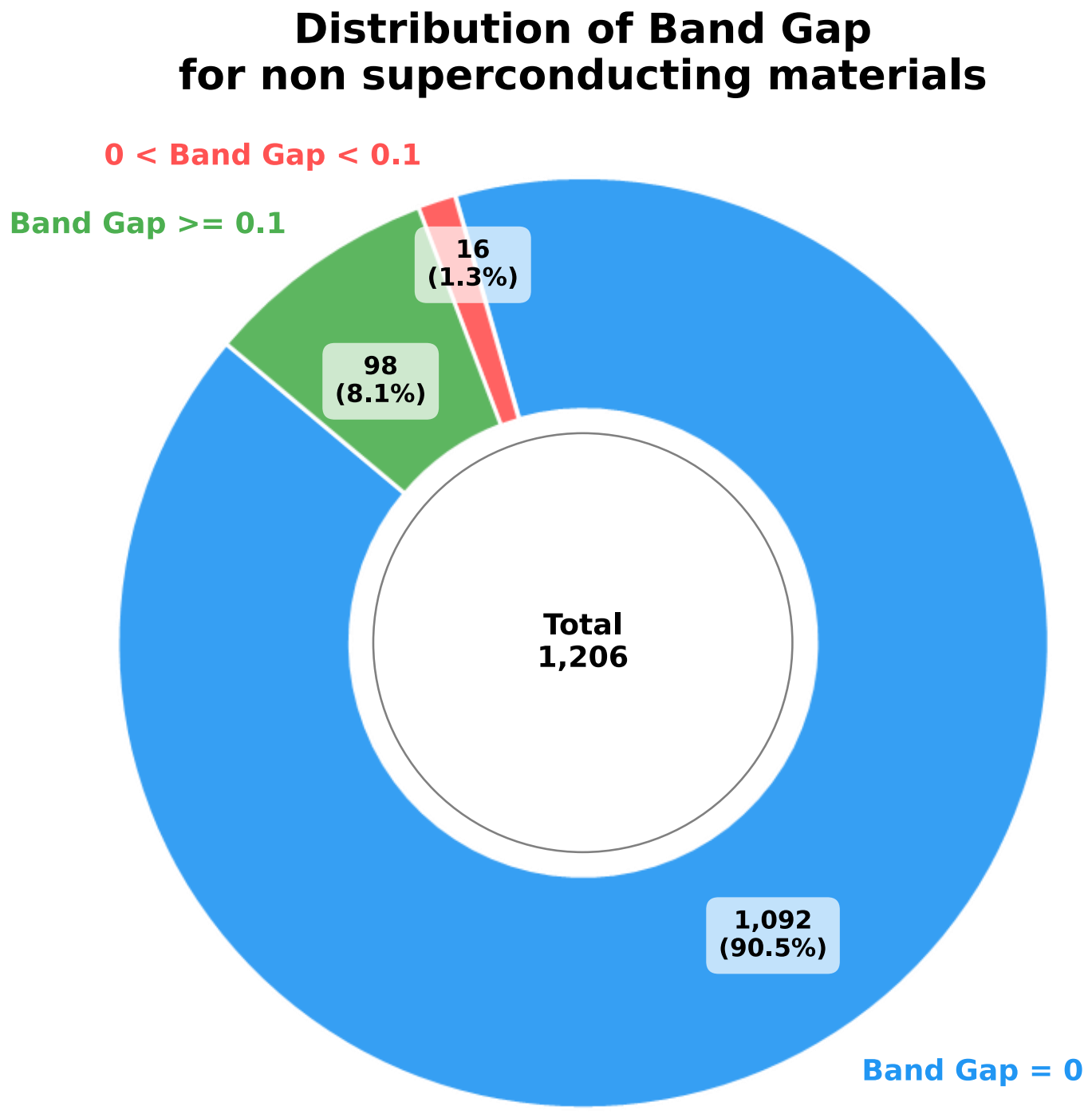

**Figure 10. Band gap distribution for 1,206 non-superconducting materials from the SuperCon dataset, overlapping with the combined Materials Project and AFlow database of 130,226 materials.**

# Supplementary information

**H. Gashmard[1], H. Shakeripour[1*], M. Alaei[1,2]**

[1] *Department of Physics, Isfahan University of Technology, Isfahan 84156-83111*
[2] *Skolkovo Institute of Science and Technology, Bolshoy Boulevard 30, bld. 1, Moscow 121205, Russia*

[*] *hshakeri@iut.ac.ir*

### Crystal Structure of Predicted Superconductors

The crystal structures of several predicted superconducting compounds incorporating Pu and H are presented below.

A preliminary structural analysis of H compounds suggests that these materials may possess discrete and molecular configurations, which may lack the extended electronic connectivity typically required for robust superconducting behavior. Despite these limitations, we propose these AI-predicted candidates for empirical validation to assess their potential as unconventional superconductors.

### A. Plutonium-Based Compounds

The first family under investigation comprises Pu-based compounds. As an example, the crystal structure of $Pu_{28}Zr$ is illustrated in Figure S1. This intermetallic compound crystallizes in a body-centered tetragonal structure with the space group $I4_1/a$. The unit cell parameters, determined from X-ray powder diffraction, are $a$ = 18.1899 Å and $c$ = 7.8576 Å at 293 K, with a unit cell volume of 2599.86 Å$^3$, reflecting the structural complexity and large size of the unit cell. Notably, the structure features a very short interatomic distance of 2.50 Å. The coordination environment of Pu atoms varies, with 12 to 14 nearest neighbors, while the presumed Zr site is coordinated by 16 neighbors. The average number of valence electrons per atom is 5.2 .

Further examination of other Pu-containing compounds reveals additional structural details. Let's have a look at the crystal structure of a few of them. For instance, $Pu_2Co_{17}$ which has predicted $T_c \approx 88.6$ K, while $Pu_3Yb$ shows $T_c \approx 82.8$ K or 72 K, depending on the phase. Similarly, $Pu_5Ru_3$, $Pu_5Os_3$, and $Pu_5Ir_3$ exhibit $T_c$ values of approximately 53.5 K.

Metallic $Pu_2Co_{17}$ crystallizes in the hexagonal $P6_3/mmc$ space group, a common structure for many intermetallic compounds. The crystal structure, detailed in Table 5 and Figure S2, features two inequivalent Pu sites: Pu1, bonded in a 12-coordinate geometry to 18 Co atoms with Pu–Co bond distances ranging from 2.93–3.27 Å, and Pu2, bonded in a 2-coordinate geometry to 20 Co atoms with Pu–Co bond distances ranging from 2.87–3.13 Å. The Co atoms occupy four inequivalent sites, forming complex polyhedral structures with Co–Co bond distances ranging from 2.36–2.68 Å .

Metallic $Pu_3Yb$ can be synthesized in either hexagonal or tetragonal crystal structures, as detailed in Table 5. The magnetic ordering differs between phases, with ferromagnetic ordering in the hexagonal phase and ferrimagnetic ordering in the tetragonal phase, as illustrated in Figure S3. Hexagonal phase has space group $P6_3/mmc$ with Lattice parameters: *a, b* = 6.84 Å, *c* = 5.66 Å, and the unit cell volume 229.25 Å$^3$, with density ρ = 13.11 g/cm$^{-3}$ . Tetragonal phase with space group I4/mmm, Lattice parameters: *a, b* = 4.66 Å, *c* = 9.46 Å, the unit cell volume is 205.71 Å$^3$ with density ρ = 14.61 g·cm$^{-3}$

Compounds of the form $Pu_5X_3$ (where X = Ru, Rh, Os, Ir, or Pt) are formed when plutonium interacts with elements from Group 8 of the periodic table. $Pu_5Ru_3$, $Pu_5Os_3$, and $Pu_5Ir_3$ are isostructural, crystallizing in the tetragonal *I4/mcm* space group. These compounds exhibit ferrimagnetic ordering, with total magnetizations of 20.02, 19.51, and 13.98 µB/f.u., respectively, as shown in Figure S4.

## B. Hydrides Compounds

The third family of interest consists of H-based compounds (see Table 4). As shown in Table 4, several our predicted hydride compounds exhibit superconductivity with a maximum $T_c \approx 100$ K at ambient pressure.

The compound $H_{1.7}C_{10}Al_{2.7}FO_{11.2}P_{2.7}$ ($C_{2826.8}H_{476}Al_{768}F_{283.8}O_{3168}P_{768}$) (Alumino fluorophosphate), identified here as one of 16 promising H-based superconducting candidates, exhibits $T_c \approx 100$ K at ambient pressure. While its hydrogen content aligns it broadly with hydride-based superconductors, the coexistence of carbon in a $C_{10}$ configuration invites speculation about unconventional bonding motifs. Notably, the limited carbon stoichiometry and heterogeneous elemental composition (Al, F, O, P) likely preclude classical fullerene- or graphene-like architectures, which require extended $sp^2$-hybridized carbon networks. Instead, the material may represent a novel hybrid phase, where localized carbon clusters interact with hydrogen and electronegative elements (F, O) to create a correlated electron system. This work warrants experimental interrogation to explore its potential for exotic quantum states. See Figure S5.

As other examples, the crystal structures of four selected hydride compounds are shown below; $NaMo_{368}H_{1410}(S_{16}O_{643})_3$ ($H_{1410}Mo_{368}NaO_{1929}S_{48}$) with predicted $T_c \approx 57$ or 71 K, $Zn_3H_{29}C_{42}N_3O_{22}$ with $T_c \approx$ 60 or 118 K, $H_{3.7}O_{6.8}U$ ($H_{944}O_{1752}U_{256}$) with $T_c \approx 33$ or 59 K, $H_{49}C_{43}$, $H_{50}C_{43}$, and $H_{49}C_{44}$ with $T_c \approx 74$ K (See Figures S6 to S9).

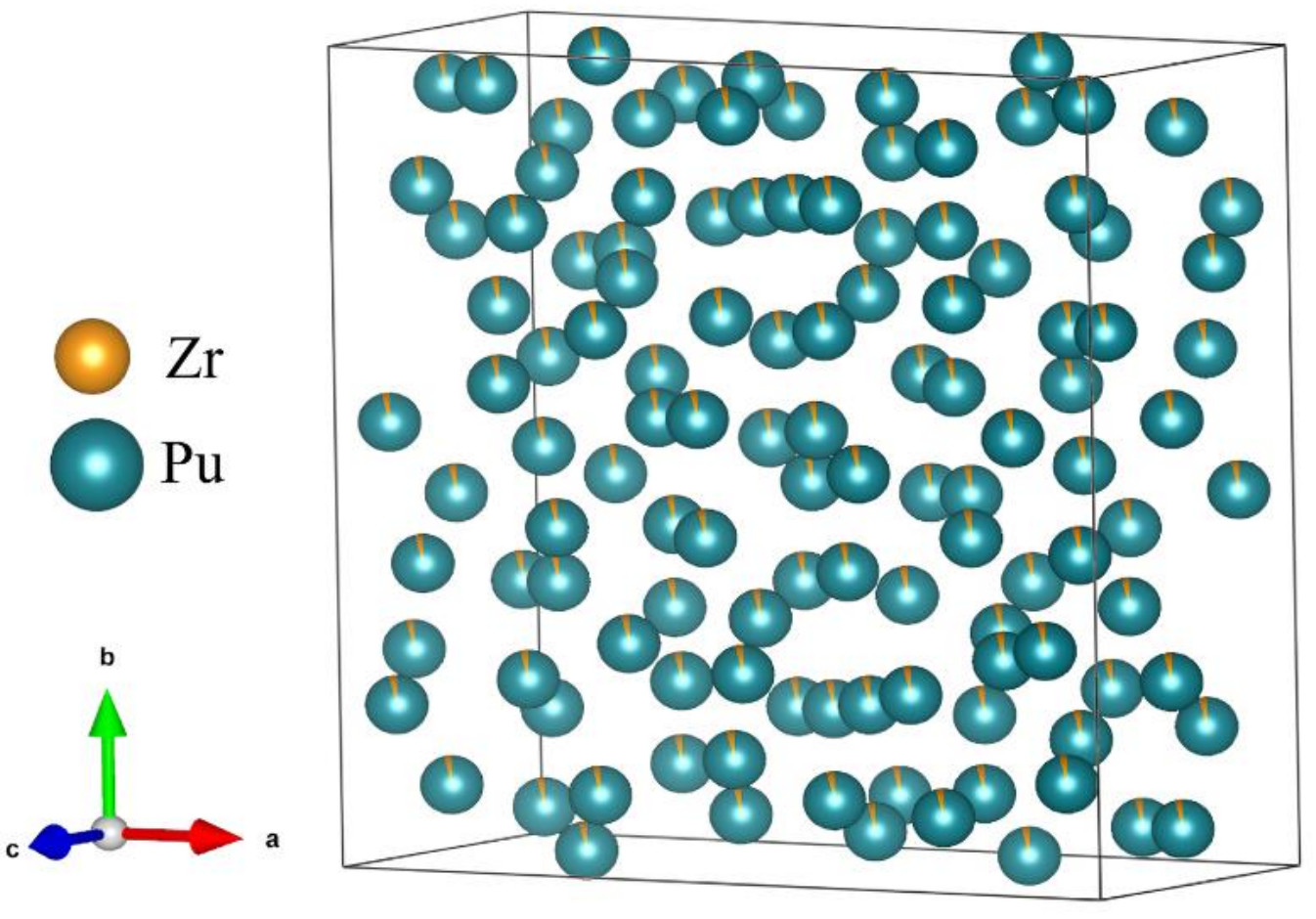

**Figure S1**. **a) The unit cell of intermetallic $Pu_{28}Zr$ compound** with predicted $T_c \approx 100$ or 96 K. **It is in the body-centered tetragonal *I*$4_1$/a space group. The lattice parameters are *a, b* =18.1899 Å, *c* = 7.8576 Å and the unit cell volume is 2599.86 Å³. Its density is ρ = 17.69 g/cm⁻³. The lines indicate the conventional cell.**

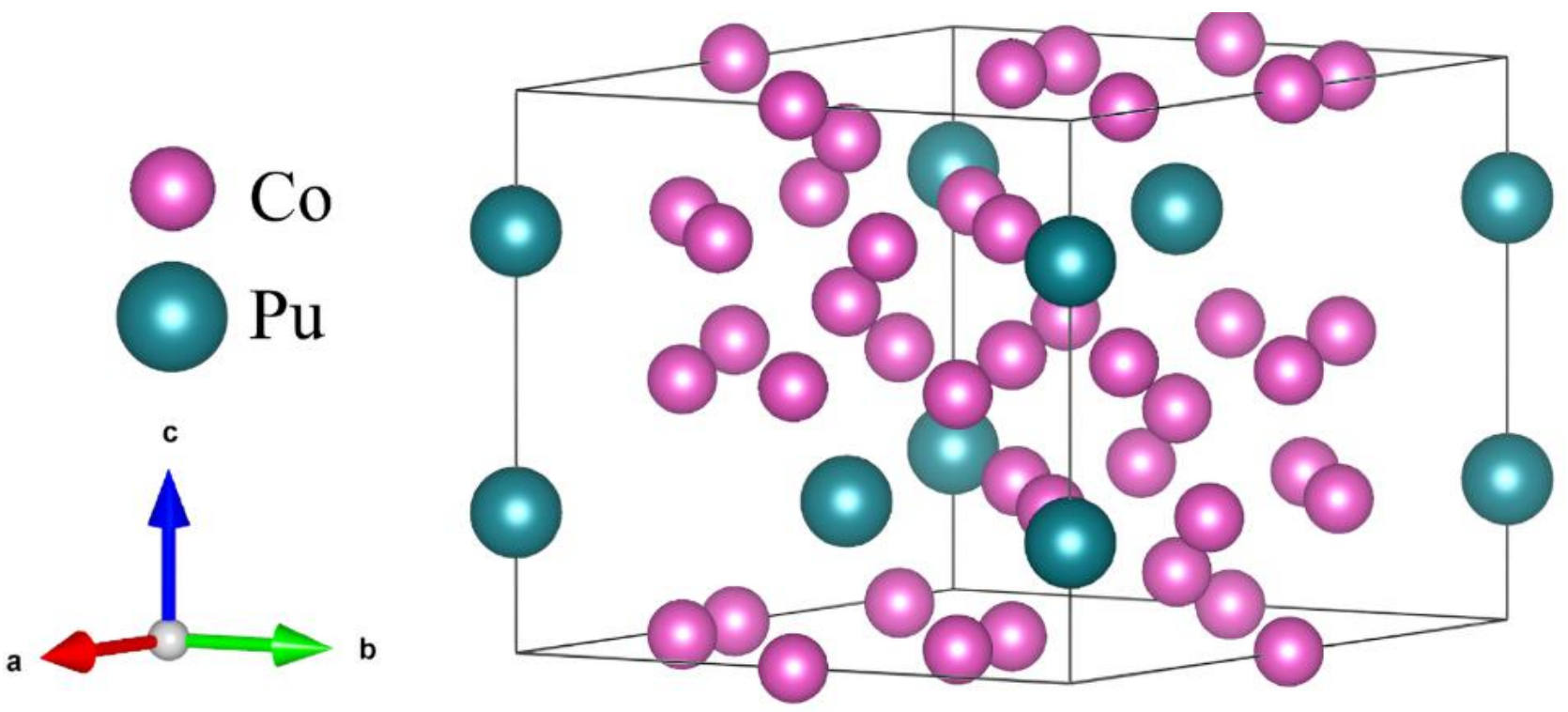

**Figure S2. The unit cell of metallic $Pu_2Co_{17}$** with predicted $T_c \approx 89$ K. **It is in the hexagonal *$P6_3/mmc$* space group. The lattice parameters are *a, b* =8.29 Å, *c* = 8.08 Å and the unit cell volume is 480.91 Å³. Its density is ρ = 10.29 g/cm⁻³.**

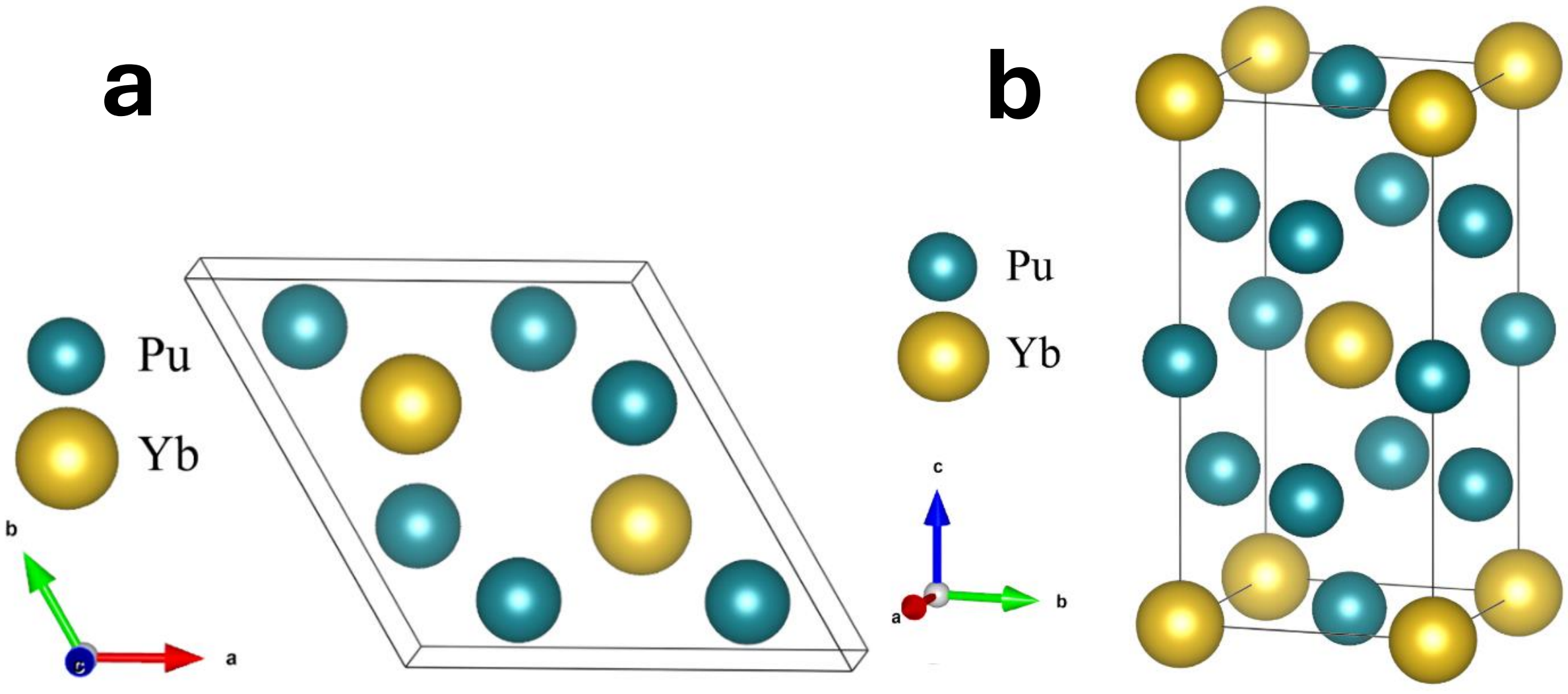

**Figure S3. The unit cell of metallic $Pu_3Yb$** with predicted $T_c \approx 83$ or 72 K**. a) Hexagonal with space group *$P6_3/mmc$*. Lattice parameters: *a, b* = 6.84 Å, *c* = 5.66 Å, the unit cell volume is 229.25 Å³, with density ρ = 13.11 g/cm⁻³ . b) Tetragonal with space group *I4/mmm*. Lattice parameters: *a, b* = 4.66 Å, *c* = 9.46 Å, the unit cell volume is 205.71 Å³ with density ρ = 14.61 g·cm⁻³.**

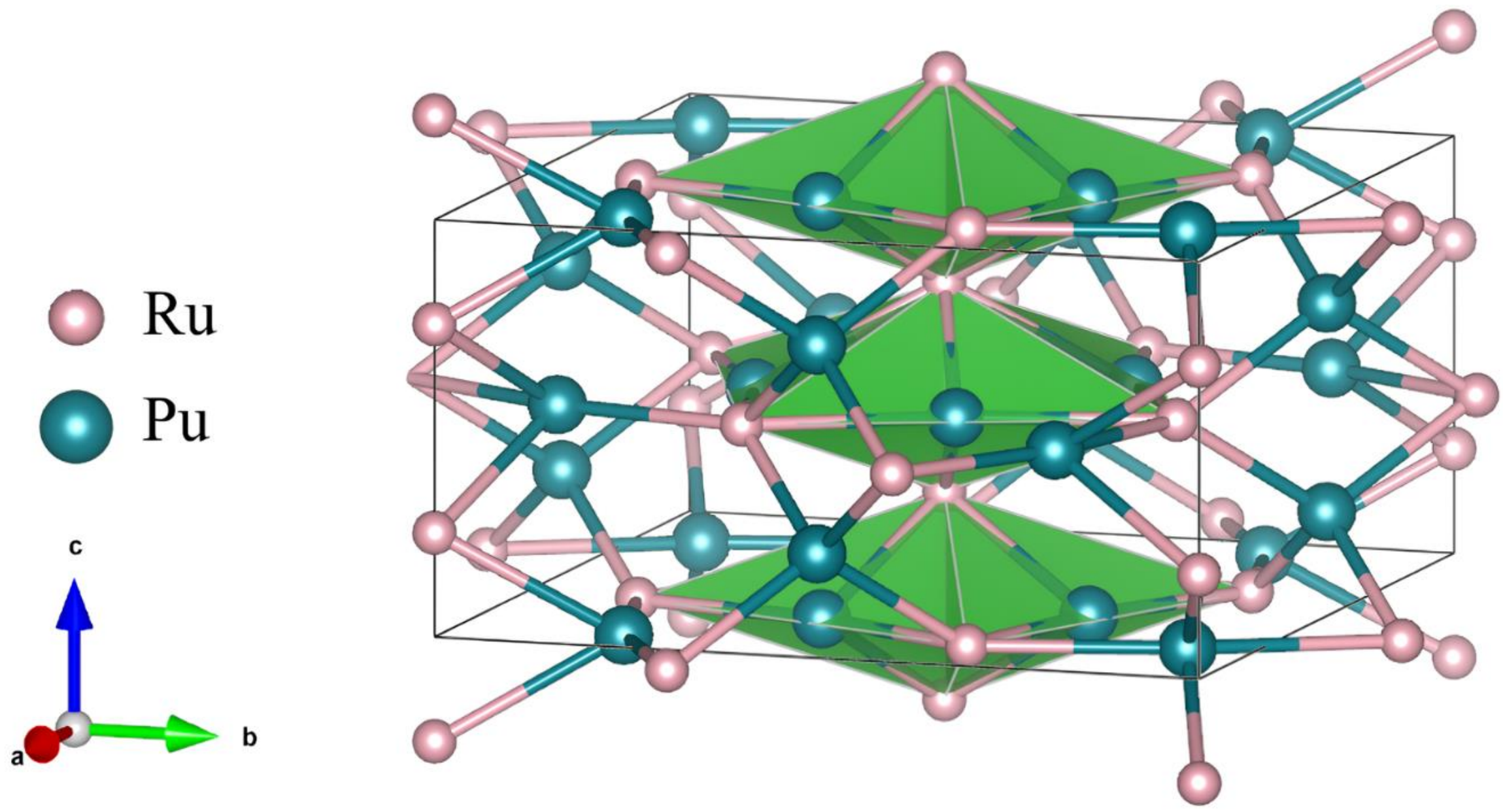

**Figure S4. The unit cell of metallic $Pu_5Ru_3$ (or $Pu_5Os_3$, and $Pu_5Ir_3$)** with predicted $T_c \approx 54$ K**. Tetragonal *I4/mcm* space group. Lattice parameters are: *a, b* = 10.82 Å, *c* = 5.67 Å, the unit cell volume is 664.05 Å$^3$, with density ρ = 15.24 g/cm$^{-3}$.**

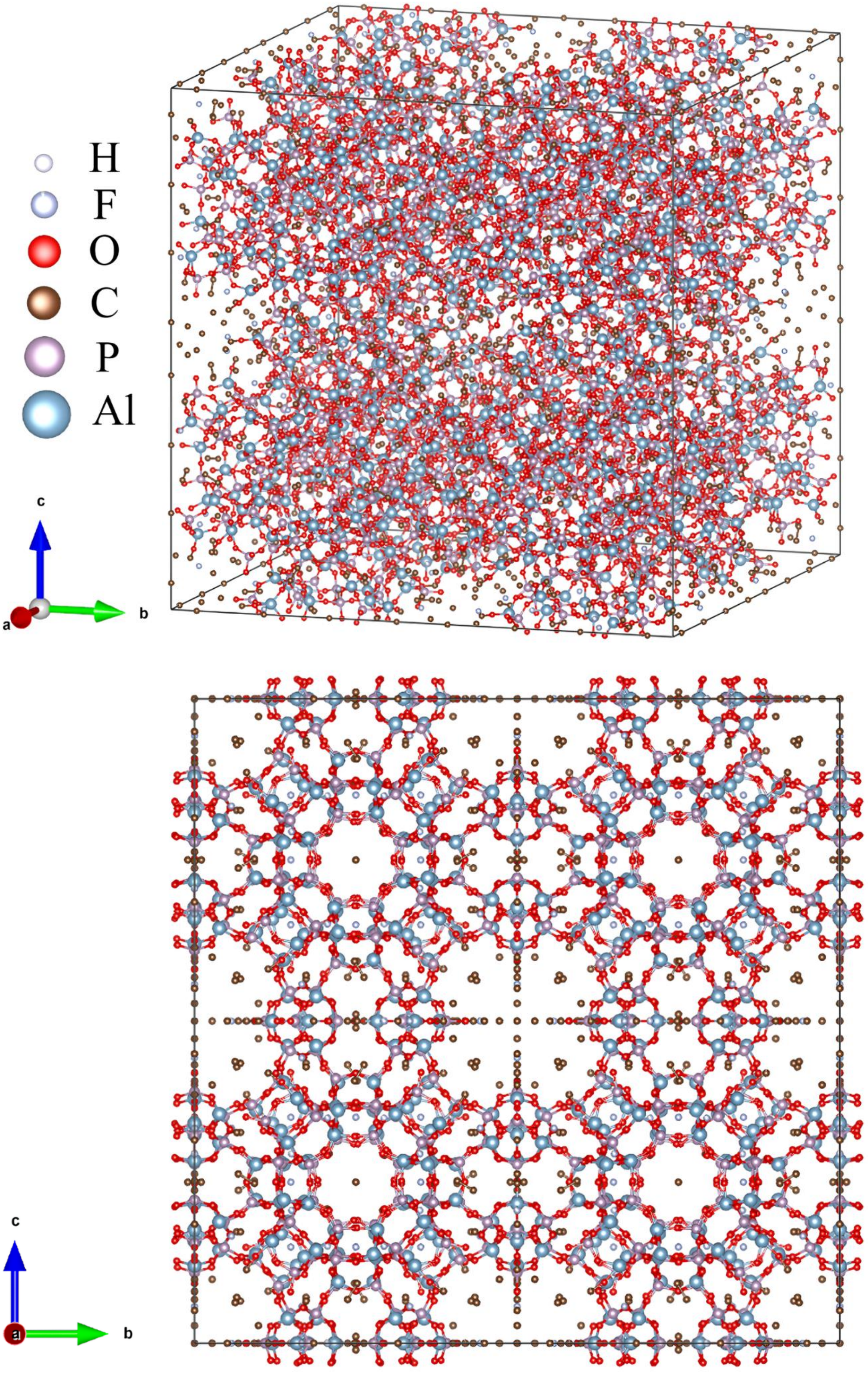

**Figure S5. The unit cell of cubic crystal structure with $Fm\overline{3}c$ space group $H_{1.7}C_{10}Al_{2.7}FO_{11.2}P_{2.7}$ ($C_{2826.8}H_{476}Al_{768}F_{283.8}O_{3168}P_{768}$)** (Alumino fluorophosphate) with a maximum predicted $T_c \approx 96$ K (or 171 K) in the Hydride-based family. **Lattice parameters are approximately: *a, b, c* = 51.3636 Å, the unit cell volume is 135508.45 Å³.**

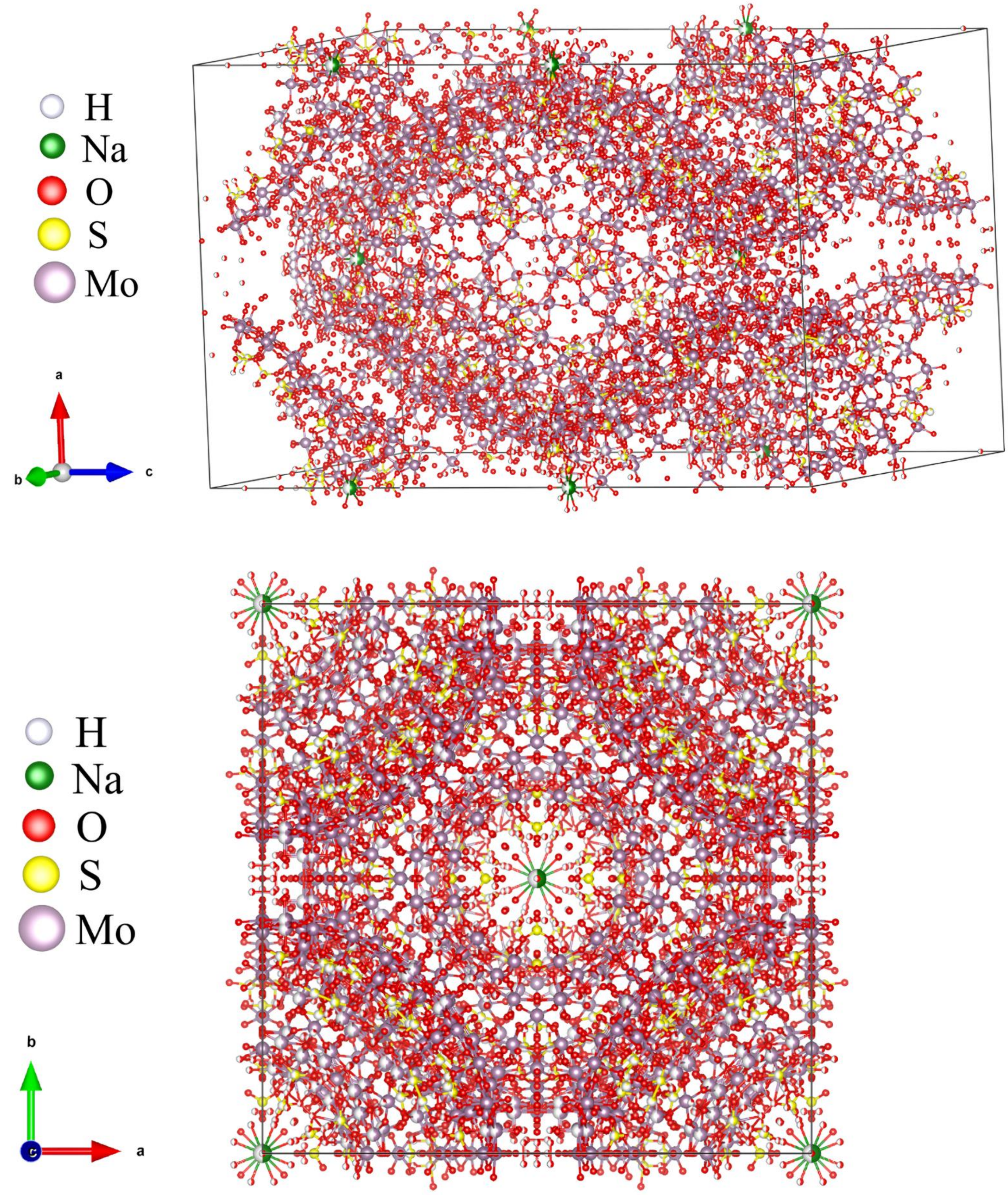

**Figure S6. The unit cell of Tetragonal crystal structure with *I*4*mm* space group $\mathbf{NaMo_{368}H_{1410}(S_{16}O_{643})_3}$ ($\mathbf{H_{1410}Mo_{368}NaO_{1929}S_{48}}$)** with predicted $T_c \approx 57$ or 71 K. **Lattice parameters are: *a, b*= 43.465 Å, *c* = 69.393 Å the unit cell volume is 131096 Å$^3$**.

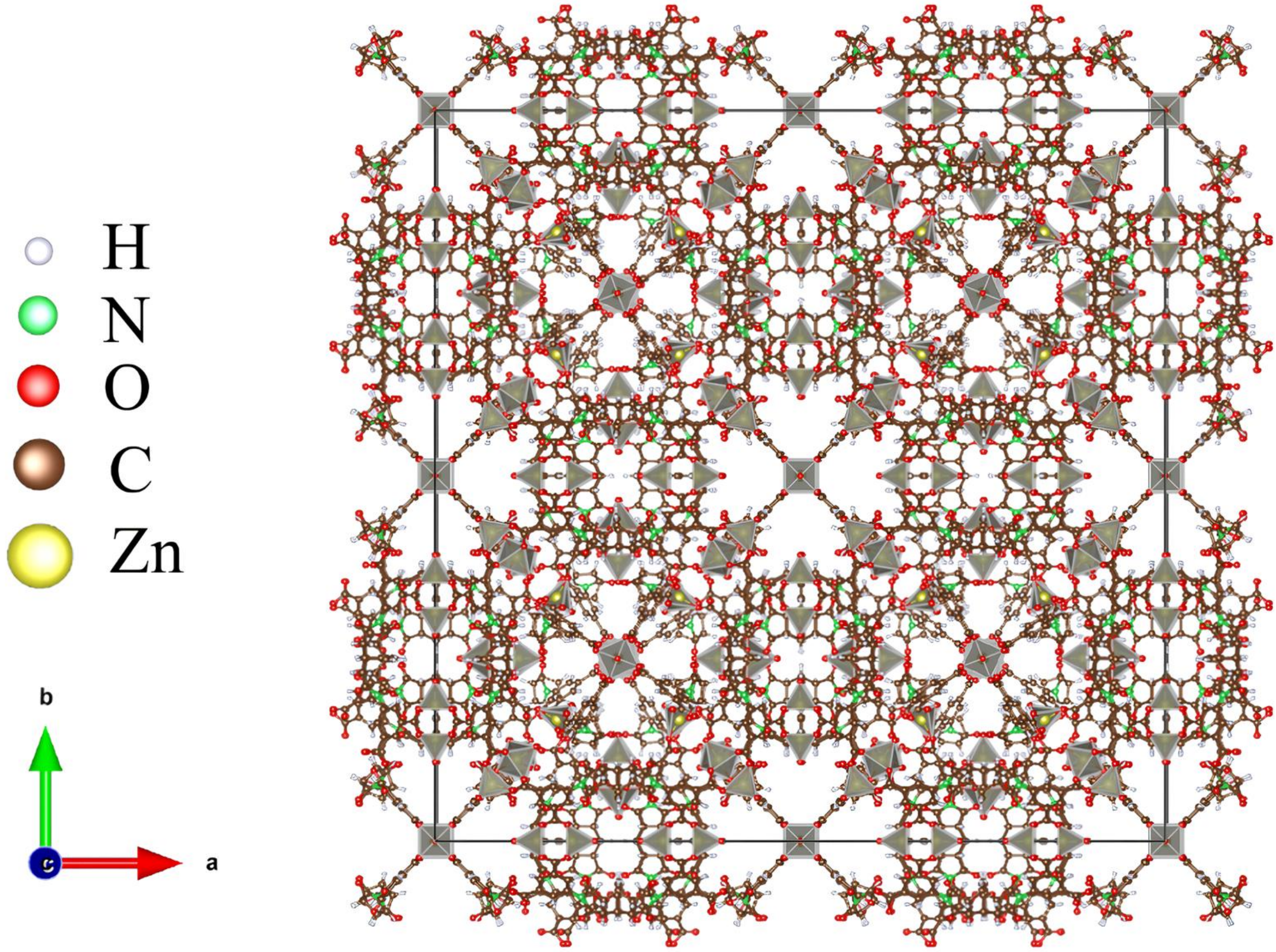

**Figure S7. The unit cell of cubic crystal structure with $Fm\overline{3}c$ space group $Zn_3H_{29}C_{42}N_3O_{22}$** with predicted $T_c \approx 60$ or 118 K. **Lattice parameters are approximately: *a, b, c* = 63.0890 Å, the unit cell volume is 251108 Å³**.

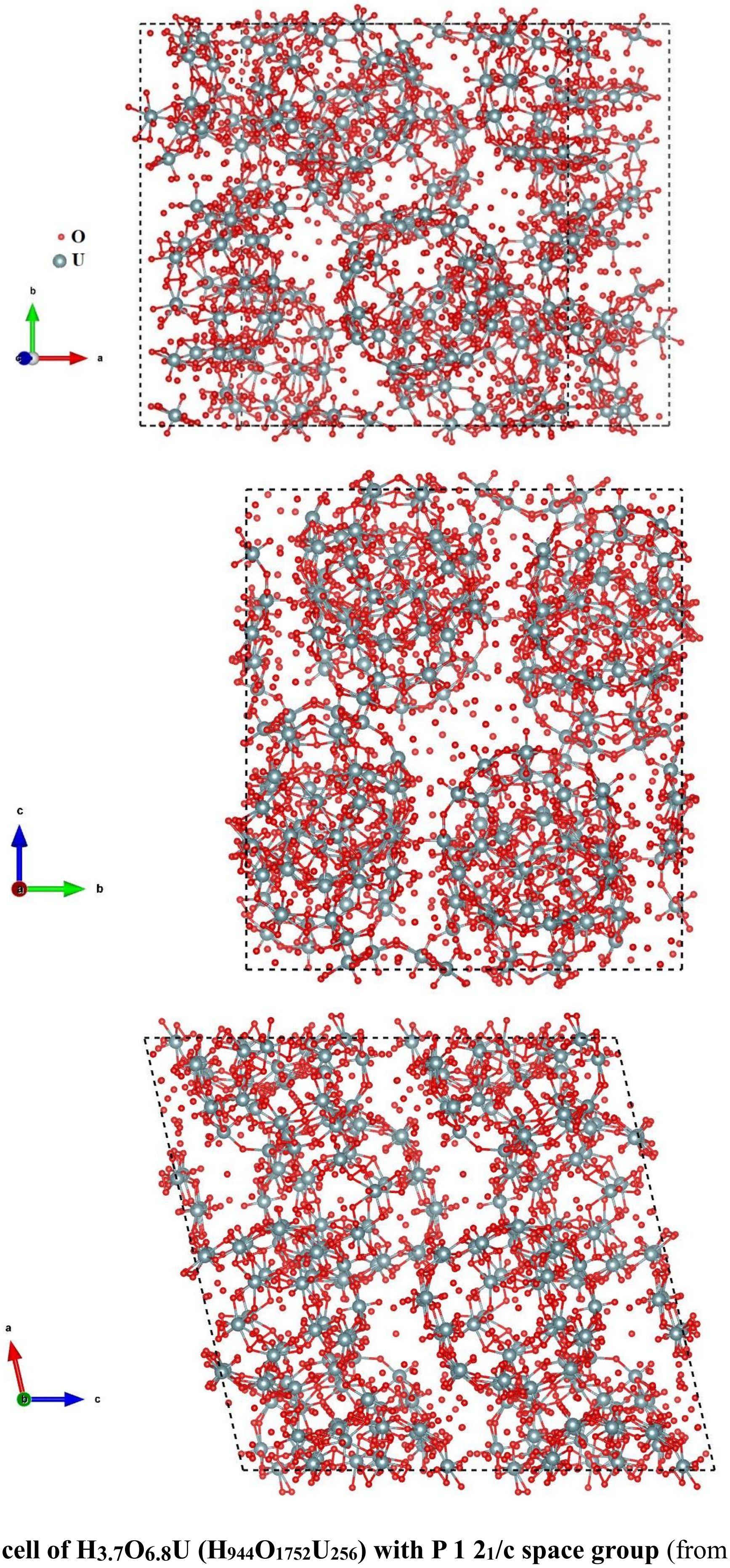

**Figure S8. The unit cell of $H_{3.7}O_{6.8}U$ ($H_{944}O_{1752}U_{256}$) with P 1 $2_1$/c space group** (from ICSD) with predicted $T_c \approx 33$ K (shown from *c*, *a* and *b* crystal directions). **Lattice parameters: *a* = 38.84 Å, *b* = 36.52 Å, *c* = 41.30 Å, the unit cell volume is 57111.35 Å$^3$, $\alpha$=90$^o$, $\beta$=102.863$^o$ $\gamma$ =90$^o$**.

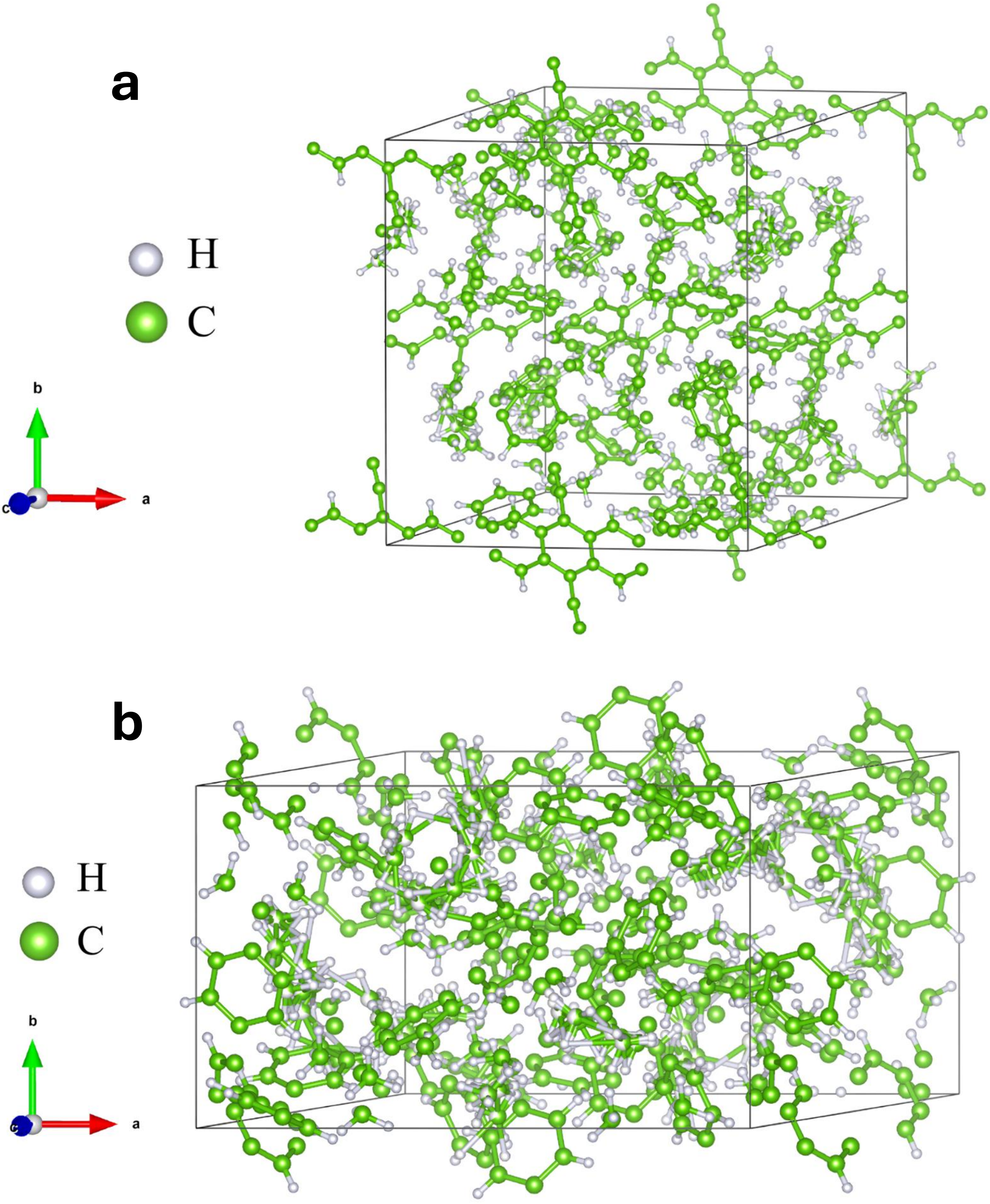

**Figure S9. a, b) The unit cell of Monoclinic $P\ 1\ 2_1/c\ 1$ space group a) $H_{49}C_{43}$ and b) $H_{50}C_{43}$** with predicted $T_c \approx 75$ and 74 K, respectively. **$H_{49}C_{43}$ has lattice parameters: *a* = 16.8988 Å, *b* = 18.7150 Å, *c* = 24.6740 Å, the unit cell volume is 7747.7 Å$^3$, $\alpha$=90$^o$, $\beta$=96.852$^o$ $\gamma$ =90$^o$**. **$H_{50}C_{43}$ has lattice parameters: *a* = 31.61 Å, *b* = 15.28 Å, *c* = 14.80 Å, the unit cell volume is 6793 Å$^3$, $\alpha$=90$^o$, $\beta$=108.29$^o$, $\gamma$ =90$^o$**.

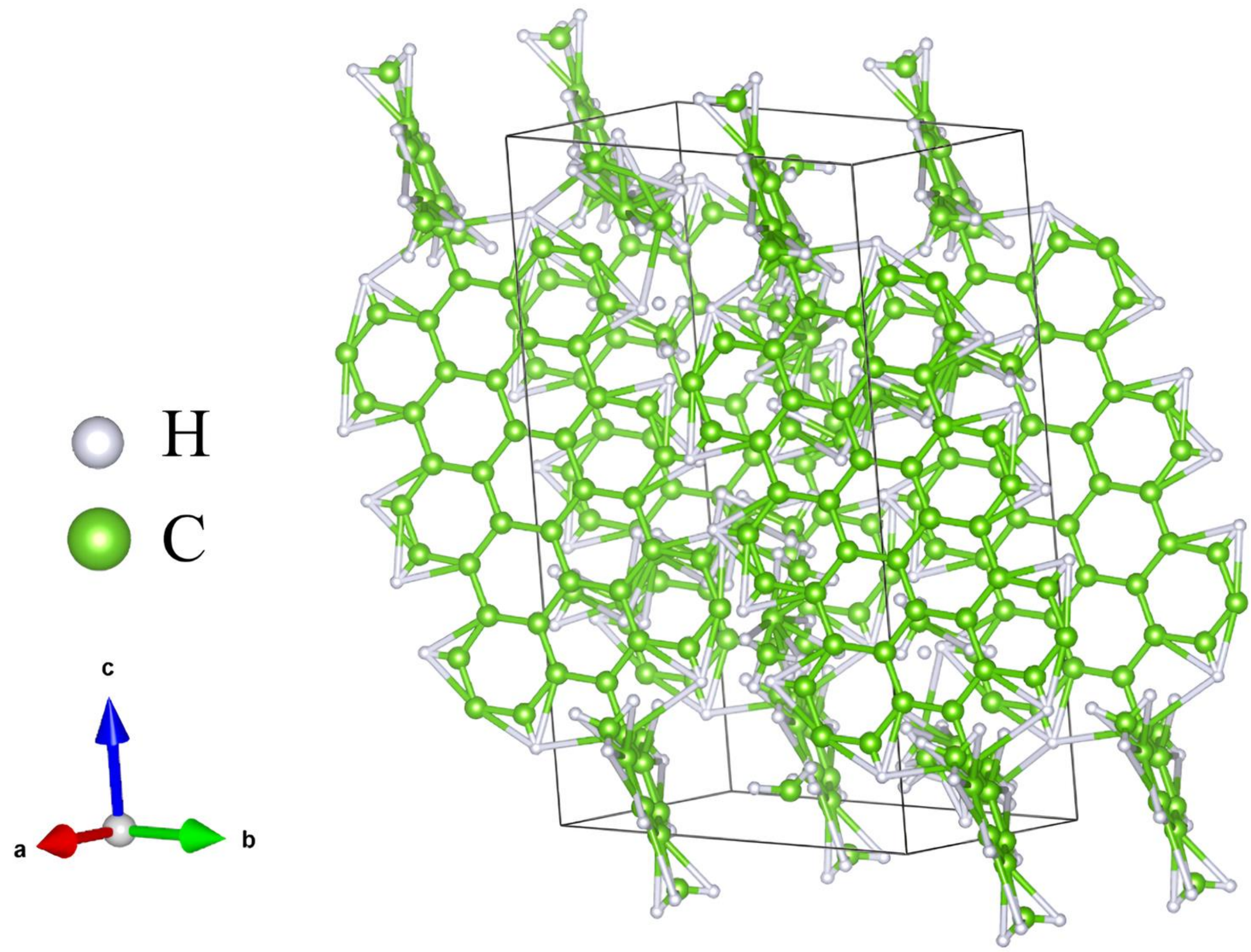

**Figure S9 c). The unit cell of crystal structure $H_{49}C_{44}$** with predicted $T_c \approx 74$ K. **It has a Triclinic structure with *P1* space group. The lattice parameters are: *a* = 7.54 Å, *b* = 12.02 Å, *c* = 19.02 Å, the unit cell volume is 1698 Å$^3$, $\alpha$=86.39$^o$, $\beta$=81.24$^o$, $\gamma$ =86.00$^o$. The lines indicate the conventional cell.**

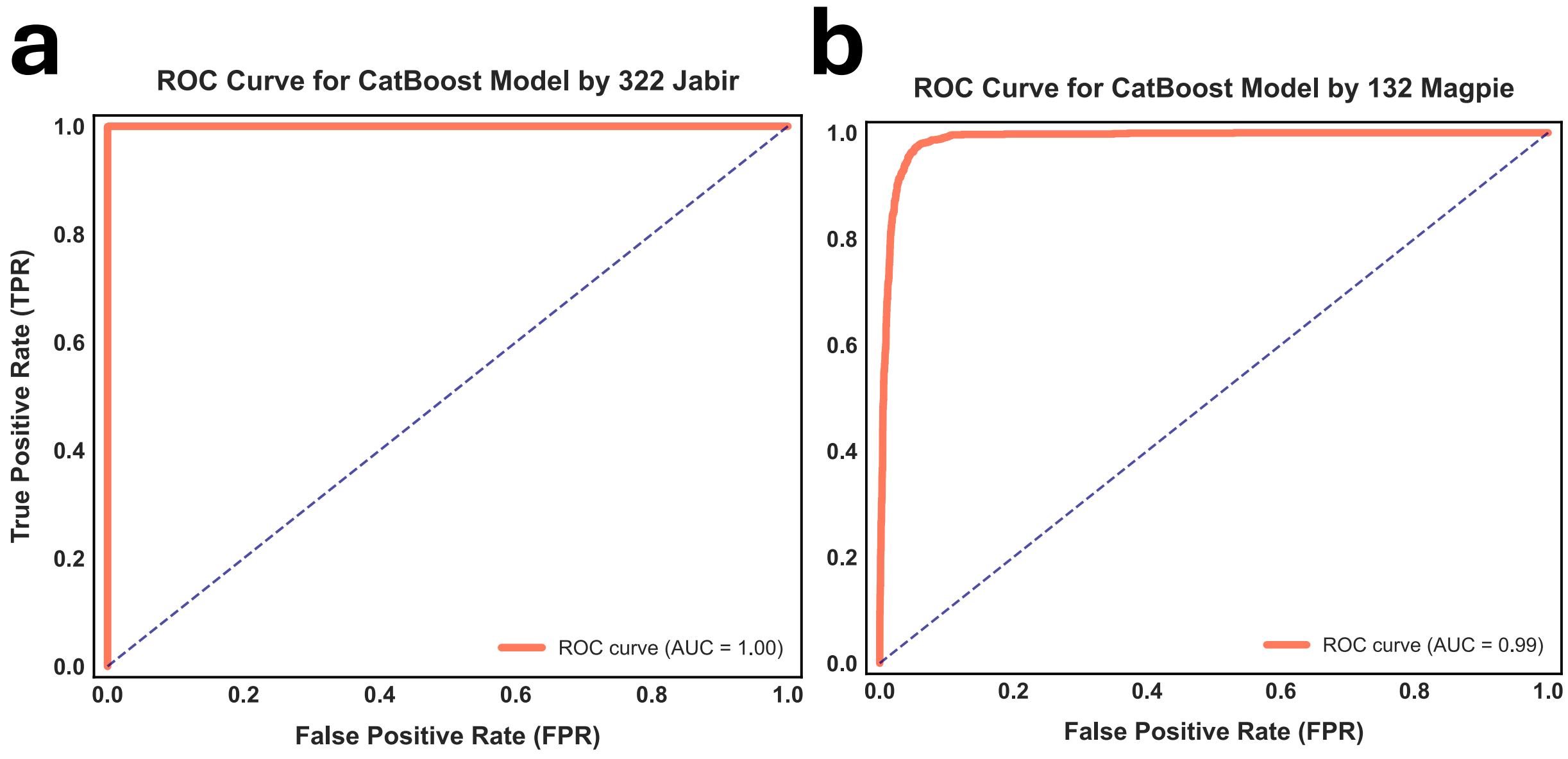

**Figure S10. Receiver Operating Characteristic (ROC) curve for the CatBoost model classification performance using features generated from a) Jabir and b) Magpie descriptors.**

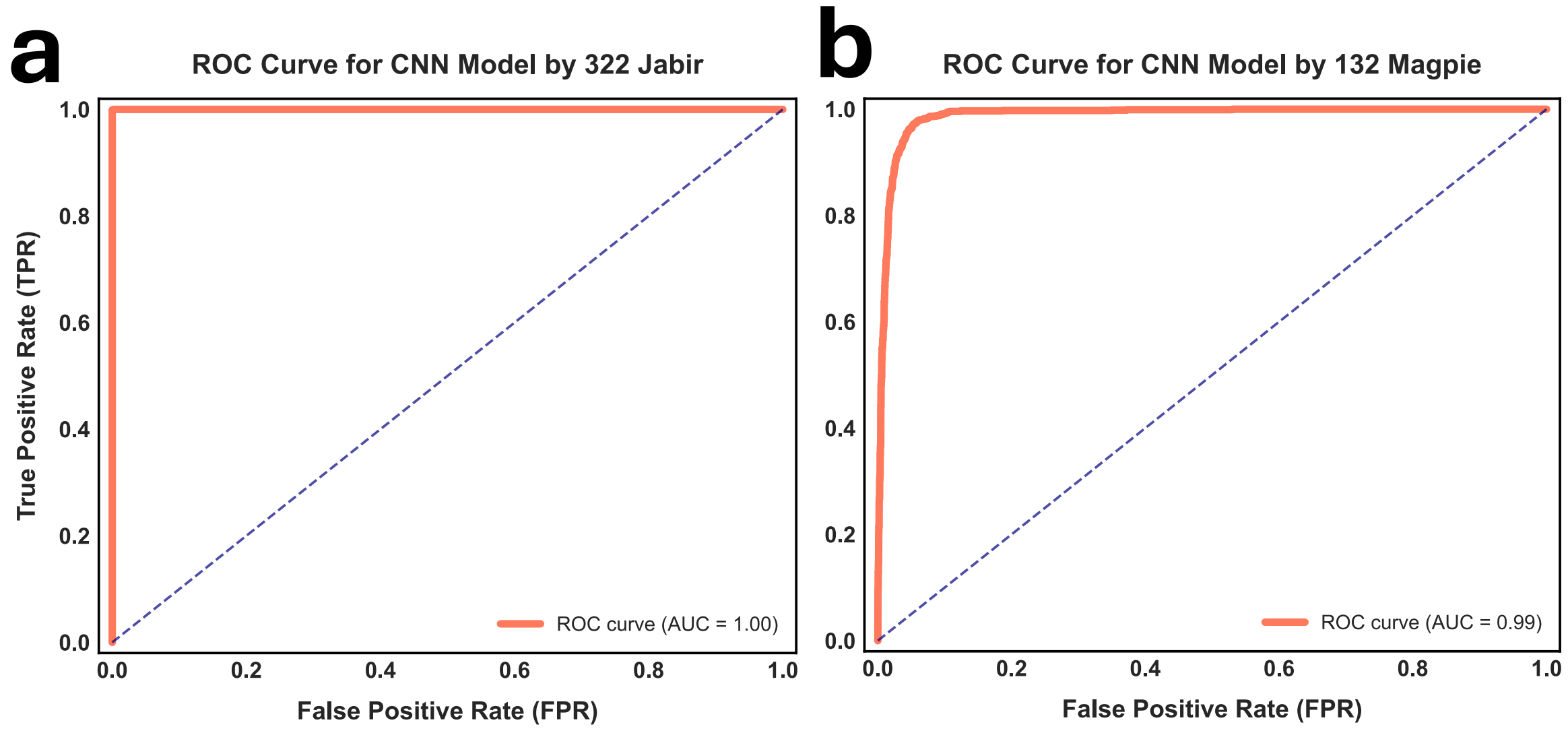

**Figure S11. Receiver Operating Characteristic (ROC) curve for the CNN model classification performance using features generated from a) Jabir and b) Magpie descriptors**.